\documentclass{article}

\PassOptionsToPackage{numbers, compress}{natbib}

    \usepackage[preprint]{neurips_2025}

\usepackage{amssymb} 
\usepackage{amsmath}
\usepackage{multirow}
\usepackage{enumitem}
\setlist[itemize]{
    leftmargin=17pt
}
\usepackage[pdftex]{graphicx}
\usepackage[utf8]{inputenc} 
\usepackage[T1]{fontenc}    
\usepackage{hyperref}       
\usepackage{url}            
\usepackage{booktabs}       
\usepackage{amsfonts}       
\usepackage{nicefrac}       
\usepackage{microtype}      
\usepackage{xcolor}         
\newcommand{\eg}{e.g.}
\newcommand{\ie}{i.e.}

\newcommand{\modelname}{MassTool}
\title{MassTool: A Multi-Task Search-Based Tool Retrieval Framework for Large Language Models}

\author{%
  Jianghao Lin$^1$, Xinyuan Wang$^1$, Xinyi Dai$^2$, Menghui Zhu$^2$, Bo Chen$^2$, \\
  \textbf{Ruiming Tang$^2$, Yong Yu$^1$, Weinan Zhang$^1$}\\
  $^1$Shanghai Jiao Tong University, $^2$Huawei Noah's Ark Lab \\\texttt{\{chiangel,wnzhang\}@sjtu.edu.cn}
}

\begin{document}

\maketitle

\begin{abstract}
Tool retrieval is a critical component in enabling large language models (LLMs) to interact effectively with external tools.  It aims to precisely filter the massive tools into a small set of candidates for the downstream tool-augmented LLMs. However, most existing approaches primarily focus on optimizing tool representations, often neglecting the importance of precise query comprehension. To address this gap, we introduce \textbf{MassTool}, a \underline{m}ulti-t\underline{as}k \underline{s}earch-based framework designed to enhance both query representation and tool retrieval accuracy. MassTool employs a two-tower architecture: a tool usage detection tower that predicts the need for function calls, and a tool retrieval tower that leverages a query-centric graph convolution network (QC-GCN) for effective query-tool matching. It also incorporates search-based user intent modeling (SUIM) to handle diverse and out-of-distribution queries, alongside an adaptive knowledge transfer (AdaKT) module for efficient multi-task learning. By jointly optimizing tool usage detection loss, list-wise retrieval loss, and contrastive regularization loss, MassTool establishes a robust dual-step sequential decision-making pipeline for precise query understanding. Extensive experiments demonstrate its effectiveness in improving retrieval accuracy. Our code is available at \url{https://github.com/wxydada/MassTool}.
\end{abstract}

\section{Introduction}
\label{sec:intro}

\begin{figure}[t]
  \centering
  \includegraphics[width=\textwidth]{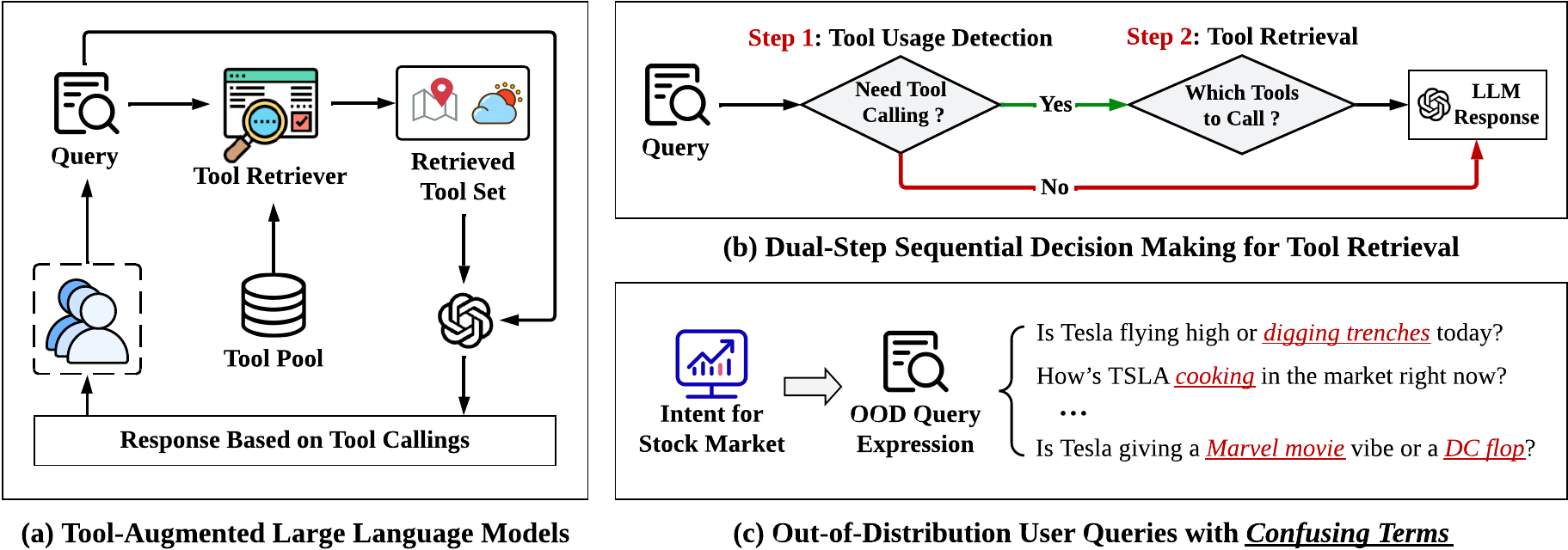}
  \vspace{-10pt}
  \caption{
  (a) The pipeline of tool-augmented large language models with tool retriever. 
  (b) The illustration of a dual-step sequential decision-making formulation of tool retrieval. 
  (c) The examples of out-of-distribution user queries with similar core user intent for stock market.
  }
  \vspace{-15pt}
  \label{fig:illustration}
\end{figure}

Recent advancements in large language models (LLMs) have significantly improved their abilities in tasks like mathematical reasoning~\citep{satpute2024can,luo2024improve,zhu2025retrieval} and natural dialogues~\citep{siro2024rethinking,joko2024doing,xi2024memocrs}. However, these models are inherently limited by their fixed parameters and static context windows~\citep{patil2023gorilla,wang2025towards}, requiring continuous retraining to keep pace with rapidly evolving real-world knowledge.
Tool learning~\citep{li2023api,parisi2022talm,qu2024tool,schick2024toolformer} offers a promising solution by allowing LLMs to dynamically access external tools. This approach enables real-time data access and complex computation, significantly enhancing LLM performance in real-world applications. However,  most previous research~\citep{lin2024hammer,shen2024hugginggpt,liu2024toolace} on tool integration primarily focuses on a small, well-defined set of APIs\footnote{In this paper, the term ``tool'' is interchangeable with ``function'' and ``API''.}. Such a formulation neglects the fact that real-world web-scale applications involve a massive number of candidate tools as it is impractical to take descriptions of all tools as the input for LLMs due to context limitations and computational constraints.  As shown in Figure~\ref{fig:illustration}(a), it is essential to develop efficient \textbf{tool retrieval} mechanisms that can quickly filter these massive tool pools into a manageable set of candidates.

Most existing tool retrieval methods leverage dense retrieval techniques~\citep{gao2024confucius,qin2023toolllm,lin2025can,yuan2024easytool} to measure semantic similarities between user queries and tool descriptions. While these approaches have advanced tool representations, \eg, capturing set-wise tool correlation~\citep{qu2024colt}, applying multi-stage refinement~\citep{zheng2024toolrerank,moon2024efficient}, and modeling diverse use cases~\citep{zhang2024data}, they often overlook the importance of precise \textbf{query representation}, which is critical for accurately capturing diverse user intents. Although related work in web search~\citep{huang2024comprehensive,lin2021graph,peng2024large,li2023agent4ranking} has extensively explored query understanding, the same level of attention has not been given to user intent modeling in tool retrieval. Previous studies~\citep{fore2024geckopt,huang2024planning,huang2023metatool,wang2024tool} have made initial attempts at query optimization but still face two major challenges:

\textit{First, tool retrieval naturally involves a \textbf{dual-step sequential decision-making} process.} As illustrated in Figure~\ref{fig:illustration}(b), this process requires determining whether a given query actually needs a tool invocation (\ie, tool usage detection) before retrieving a set of suitable tools (\ie, tool retrieval). Many queries can be handled directly by LLMs without external tools, \eg, chit-chat, text summarization, and general knowledge responses. However, existing research on tool retrieval usually overlooks the crucial preliminary step and assumes that all training queries are tool-dependent, which can lead to suboptimal performance, as it causes a biased view of the input queries and global user intents.

\textit{Secondly, user queries often exhibit significant \textbf{out-of-distribution (OOD)} variation, with diverse and overlapping intents expressed in varying forms.}
As shown in Figure~\ref{fig:illustration}(c), even when targeting similar intents, such as stock market information, the phrasing of these queries can differ greatly due to factors like regional dialects and cultural nuances~\cite{joshi2024natural}.
This diversity poses a major challenge for effective intent modeling, especially in tool retrieval, since the retriever is often a smaller language model (\eg, BERT-base with 110M parameters) to ensure lightweight integration with downstream LLMs. 

To address these challenges, we introduce \textbf{MassTool}, a \underline{m}ulti-t\underline{a}sk \underline{s}earch-ba\underline{s}ed \underline{tool} retrieval framework designed to optimize query representation under a joint modeling of both tool usage detection and tool retrieval tasks with a two-tower architecture. The \textit{tool usage detection} tower predicts the probability that a query requires external tools and injects the hidden states as detective knowledge to the \textit{tool retrieval} tower. The \textit{tool retrieval} tower includes a \textbf{query-centric graph convolution network} (QC-GCN) to capture collaborative query-tool patterns, a \textbf{search-based user intent modeling} (SUIM) module for handling diverse and OOD queries by integrating semantically relevant neighbors with a dynamic filtering mechanism, and an \textbf{adaptive knowledge transfer} (AdaKT) module for integrating knowledge across the two towers. Finally, we fuse the enhanced query representations for query-tool matching score estimation. In this way, we establish a dual-step sequential decision-making pipeline to optimize query representations for precise tool retrieval.  

Moreover, for the preliminary step of tool usage detection, we construct a new tool usage detection dataset (\textbf{ToolDet}) with 31,229 non-tool-invocation queries. This dataset is built by filtering the Chatbot Arena dataset~\cite{zheng2023judging} and validated by human annotators and LLMs. Our tool usage detection dataset is \textbf{dataset-agnostic} and can serve as an auxiliary resource for existing tool retrieval datasets. 

The main contributions of this paper are as follows:
\begin{itemize}[leftmargin=10pt]
    \item We are the first to introduce the multi-task modeling paradigm in tool retrieval for large language models, integrating both tool usage detection and tool retrieval to improve query understanding.
    \item We introduce MassTool, a two-tower architecture featuring three core modules for query representation optimization: QC-GCN for collaborative query-tool pattern learning, SUIM for handling diverse and OOD queries, and AdaKT for cross-task knowledge fusion.
    \item We release the first dataset for tool usage detection (\ie, ToolDet)  consisting of queries that do not require tool invocations, enabling multi-task training for improved retrieval performance. The dataset is open-sourced to support future research.
    \item Extensive experiments on three public datasets validate the superiority of our proposed MassTool for query comprehension and tool retrieval, compared with existing baseline methods.
\end{itemize}

\section{Related Work}
\label{sec: related work}

\paragraph{Tool-Augmented Large Language Models.}
Integrating external tools with large language models (LLMs) has emerged as a key direction for enhancing their capabilities in complex tasks~\cite{qu2024tool,parisi2022talm}. 
Existing approaches can be broadly classified into \textit{tuning-free} and \textit{tuning-based} methods. 
Tuning-free methods~\cite{liu2024summary,shen2024hugginggpt,yao2022react} rely on in-context learning with prompt engineering techniques, such as chain-of-thought~\cite{wei2022chain} prompting and demonstration ordering~\cite{liu2024demorank}, to guide LLMs in complex reasoning~\cite{liu2024summary,shen2024hugginggpt,yao2022react}. 
Tuning-based methods~\cite{gao2024confucius,qin2023toolllm,schick2024toolformer,tang2023toolalpaca,xu2023tool,liu2024toolace} directly finetune LLMs on specialized tool datasets, significantly enhancing their function-calling abilities through supervised training. 
Real-world applications require LLMs to handle thousands of potential tools across diverse domains. Given context and latency constraints, the visible candidate set is typically small, highlighting the need for efficient tool retrieval to filter these vast tool pools into a manageable set~\cite{lin2024hammer,liu2024toolace,liu2024apigen}.

\paragraph{Tool Retrieval.}
Tool retrieval is critical for identifying relevant tools from large repositories in response to user queries. Methods generally fall into two categories: \textbf{term-based} and \textbf{semantic-based} approaches. Term-based methods, like TF-IDF~\cite{sparck1972statistical} and BM25~\cite{robertson2009probabilistic}, rely on sparse lexical matching, while semantic-based methods leverage dense representations to capture deeper query-tool relationships, like ANCE~\cite{xiong2020approximate}, TAS-B~\cite{hofstatter2021efficiently}, coCondensor~\cite{gao2021unsupervised}, and Contriever~\cite{izacard2021unsupervised}. 
Recent work has focused on adapting deep semantic models specifically for tool retrieval, incorporating task-specific architectures and training strategies to improve relevance ranking~\cite{zheng2024toolrerank,mu2024adaptive,moon2024efficient,zhang2024data,qu2024colt,gao2024ptr}. For example, 
COLT~\cite{qu2024colt} conducts scene-aware collaborative learning for better tool representations and proposes a new metric to assess the tool retrieval performance based on completeness.

\section{Problem Formulation}
\label{sec:preliminary}

Suppose that we have a global tool set \(\mathcal{T}=\{(t_i,d_i)\}_{i=1}^{T}\), where the \(i\)-th tool \(t_i\) is associated with its description \(d_i\). 
Each instance of the tool retrieval dataset \(\mathcal{D}\) can be formulated as a triplet \((q,y,\mathcal{T}_q)\), where \(q\) is the query, \(y\in\{1,0\}\) is the \textit{tool usage label} indicating whether the query needs tool invocation or not, and \(\mathcal{T}_q\) is the golden tool set required by the query \(q\). 
If query \(q\) does not require tool invocation (\ie, \(y=0\)), the golden tool set would be empty (\ie, \(\mathcal{T}_q=\varnothing\)). 
Therefore, the dataset \(\mathcal{D}\) can be divided into two subsets:
\begin{equation}
\begin{aligned}
\mathcal{D}&=\mathcal{D}_{det}\cup\mathcal{D}_{ret}, \\
    \mathcal{D}_{det}&=\{(q,y,\mathcal{T}_q)\in\mathcal{D}\;|\;y=0,\mathcal{T}_q=\varnothing\},\\
    \mathcal{D}_{ret}&=\{(q,y,\mathcal{T}_q)\in\mathcal{D}\;|\;y=1,\mathcal{T}_q \ne \varnothing\}.
    \label{eq:dataset}
\end{aligned}
\end{equation}
Next, we can define the tool retrieval task in our paper as a \textit{dual-step sequential decision-making problem}. 
That is, given the user query \(q\), our goal is to perform the following two tasks:
\begin{itemize}[leftmargin=10pt]
    \item \textbf{Tool Usage Detection}. We should first predict whether query \(q\) truly needs tool invocation for tool-augmented LLMs or not, \ie, the tool usage label \(y\).
    \item \textbf{Tool Retrieval}. If the tool usage label is predicted as positive, we should then retrieve the top-\(K\) most suitable tools from the global tool pool, \ie, the golden tool set \(\mathcal{T}_q\subset\mathcal{T}\).
\end{itemize}
It is worth noting that previous works on tool retrieval generally assume that all the queries are tool-dependent, \ie, the dataset they use is merely \(\mathcal{D}_{ret}\) in Eq.~\ref{eq:dataset}. 
In contrast, we are the first to formulate the tool retrieval task as a dual-step sequential decision-making problem, and further extend the dataset \(\mathcal{D}_{ret}\) with \(\mathcal{D}_{det}\). 
In this way, we can conduct the multi-task joint modeling, and incorporate the information gain between the two predictive stages to further enhance the retrieval performance.

\section{Methodology}
\label{sec:method}



\begin{figure*}[t]
  \centering
  \includegraphics[width=\textwidth]{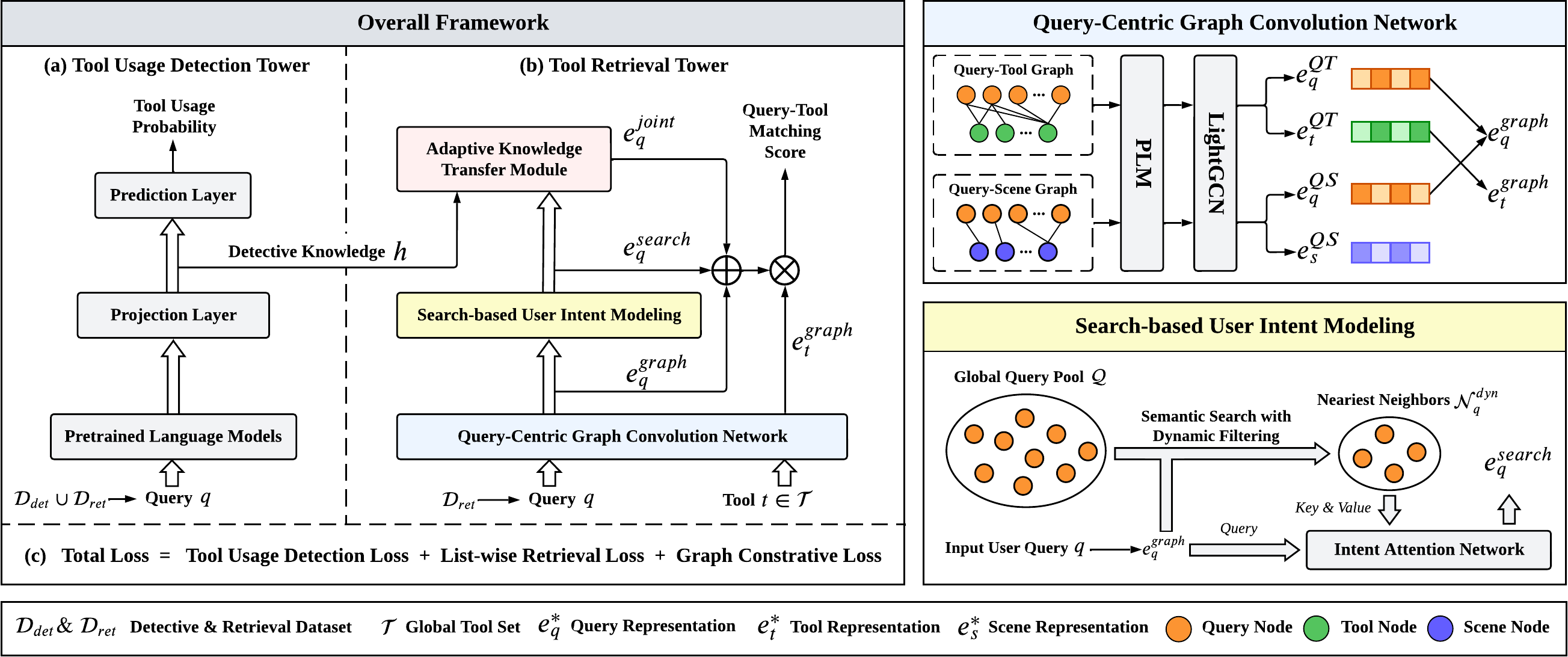}
  \vspace{-10pt}
  \caption{
  The overall framework of MassTool.
  }
  \vspace{-10pt}
  \label{fig:framework}
\end{figure*}

As shown in Figure~\ref{fig:framework}, we formulate a dual-step sequential decision-making problem, and perform a joint modeling of both tool usage detection and tool retrieval. 
The MassTool framework follows the two-tower architecture consisting of (a) tool usage detection tower, and (b) tool retrieval tower. 



\subsection{Tool Usage Detection}
\label{sec:tool detection}

The tool usage detection tower focuses on determining whether tool invocation is required for a given query. 
As shown in Figure~\ref{fig:framework}, we input the query \(q\in\mathcal{D}\) through the pretrained language model (PLM) and projection layer to obtain the detective knowledge \(h\). 
The projection is a linear layer with ReLU activation~\cite{agarap2018deep}.
The detective knowledge \(h\) encodes the semantic and contextual features essential for tool usage detection, and is then fed into the final prediction layer for tool usage probability estimation, followed by binary cross-entropy loss:
\begin{equation}
    \hat{y} = \sigma\left(\operatorname{Linear}(h)\right) \in (0, 1), 
\end{equation}
\begin{equation}
    \mathcal{L}_{det} = y \log \hat{y} + (1-y)\log(1-\hat{y}),
    \label{eq:tool detection loss}
\end{equation}
where \(\sigma(\cdot)\) is the sigmoid function, and \(y\in\{0,1\}\) is the label.
Moreover, the detective knowledge \(h\) will be passed to the adaptive knowledge transfer module in the tool retrieval tower. 

\subsection{Tool Retrieval}

The tool retrieval aims to estimate the matching score between the query \(q\) and each tool \(t\) from the global tool set \(\mathcal{T}\). 
Then, we can sort the tools based on the estimated scores and select the top-ranked ones for downstream LLMs.
As shown in Figure~\ref{fig:framework}, the tool retrieval tower comprises three core components: query-centric graph convolution network (QC-GCN), search-based user intent modeling (SUIM), and adaptive knowledge transfer module (AdaKT).

\subsubsection{Query-Centric Graph Convolution Network (QC-GCN)}
\label{sec:QC-GCN}

To better model the relational knowledge between queries and tools, as well as the contextual information for application scenes, we regard the queries and tools as nodes, and construct the following two bipartite graphs with a centric view towards the queries:
\begin{itemize}[leftmargin=10pt]
    \item The \textit{\textbf{query-tool graph}} links a query \(q\) to each tool \(t\) in its golden tool set \(\mathcal{T}_q\). 
    The golden-tool linkages for different queries can overlap with each other, thereby conveying the contextual connections and collaborative patterns among queries and tools.
    \item The \textit{\textbf{query-scene graph}} is built to represent the global patterns by defining the golden tool set \(\mathcal{T}_q\) for a given query as a unified scene  \(s\), following previous work~\cite{qu2024colt}. 
    The tools in one golden tool set \(\mathcal{T}_q\) are jointly required to fully address one user request. For example, to make a travel plan for the user, LLMs have to incorporate tools like weather forecasts, hotel booking, restaurant reservations, etc. 
    The joint modeling of scene-level information can help capture the intent-level correlation, and deepen the insights for user query understanding.
\end{itemize}
After constructing the two bipartite graphs, we apply a query-centric graph convolution network using LightGCN~\cite{he2020lightgcn} to capture high-order collaborative signals among queries, tools, and scenes, resulting in final representations for the query-tool graph: \(e_q^{QT}\) and \(e_t^{QT}\), where QT stands for query-tool graph.
Due to the page limitation, details about lightGCN can be found in Appendix~\ref{app:QC-GCN}.
We can conduct similar operations on the query-scene graph, resulting in representations $e_q^{QS}$ and $e_s^{QS}$. 
Finally, we calculate the graph-enhanced representations for query $q$ and tool $t$:
\begin{equation}
    e_q^{graph}=e_q^{QT}+e_q^{QS},\;\; e_t^{graph}=e_t^{QT}.
\end{equation}
As shown in Figure~\ref{fig:framework}, the tool representation $e_t^{graph}$ is preserved for the final matching score estimation, and the query representation $e_q^{graph}$ would be further enhanced via the search-based user intent modeling and adaptive knowledge transfer module.

\subsubsection{Search-based User Intent Modeling (SUIM)}

The user queries are often out-of-distribution (OOD) with diverse and overlapping intents in varying forms, making it difficult to directly match them to the correct tools. 
To this end, as shown in Figure~\ref{fig:framework}, we propose search-based user intent modeling to further enrich the query representation based on its semantically nearest neighbor queries. 

To be specific, we maintain a global query pool based on the retrieval training dataset $\mathcal{Q}=\{q|(q,y,\mathcal{T}_q)\in\mathcal{D}_{ret}\}$. 
Then, we first conduct semantic search to retrieve the top-$K$ semantically relevant queries based on cosine similarity. By sorting the cosine similarity, we can get the top-$K$ nearest neighbor set $\mathcal{N}_q^K$ for the target query $q$.

While semantic search retrieves the most relevant neighbors with potentially similar user intents, some rare or unconventional queries may still result in the retrieval of irrelevant neighbors. To address this, we apply dynamic filtering to filter the top-$K$ neighbor set $\mathcal{N}_q^{K}$ with a hyperparameter threshold $\epsilon$, resulting in the final neighbor set
\(    \mathcal{N}_q^{dyn} = \left\{q^{\prime}|s_{cos}(q,q^{\prime})>\epsilon,q^{\prime}\in\mathcal{N}_q^{K}\right\}.\)

After obtaining the dynamically filtered nearest neighbors, we design the intent attention network to fuse the embedded knowledge into the search-enhanced query representations:
\begin{equation}
\begin{aligned}
    \alpha_{q^{\prime}} &= \frac{\exp(e_{q}^{graph} \,W \,e_{q^{\prime}}^{graph})}{\sum\nolimits_{q^{\prime\prime}\in\mathcal{N}_q^{dyn}} \exp(e_{q}^{graph} \,W \,e_{q^{\prime\prime}}^{graph})}, \\
    e_q^{search} &= \sum\nolimits_{q^{\prime}\in\mathcal{N}_q^{dyn}} \;\,\alpha_{q^{\prime}} \, \times \, e_{q^{\prime}}^{graph},
\end{aligned}
\end{equation}
where $W$ is a learnable matrix to capture the attentive correlation between queries. 
In this way, the SUIM module can handle diverse query distributions and overlapping intents, producing robust search-enhanced query representations that facilitate effective tool retrieval, even under OOD scenarios.

In practice, the global query pool $\mathcal{Q}$ can be dynamically expanded with the oncoming user interactive data, which makes sure that we can progressively tackle the ever-evolving and diverse user queries with such search-based user intent modeling techniques.

\subsubsection{Adaptive Knowledge Transfer Module (AdaKT)}
\label{sec:knowledge transfer}

The adaptive knowledge transfer module is designed to incorporate the search-enhanced query representation $e_q^{search}$ with the detective knowledge $h$ produced by the tool usage detection tower. 
We implement the transfer function with the element-wise gating mechanism for the jointly fused query representation:
\(e_q^{joint}=\sigma\left(\operatorname{Linear(h)}\right)\otimes e_q^{search},
\)
where $\sigma(\cdot)$ is the sigmoid function, and $\otimes$ is the element-wise product operator. 
We also provide the empirical study in Section~\ref{sec:transfer func exp} about the different choices of the transfer function, demonstrating that this gating mechanism outperforms alternative fusion strategies such as attention, concatenation, and addition.

As shown in Figure~\ref{fig:framework}, with the three core components introduced above (\ie, QC-GCN, SUIM, and AdaKT), we obtain the graph-enhanced query representation $e_q^{graph}$, the search-enhanced query representation $e_q^{search}$, and the jointly fused query presentation $e_q^{joint}$. 
Hence, we normalize these three query representations, add them together, and estimate the matching score estimation with the normalized graph-enhanced tool representation $e_t^{graph}$: 
\begin{equation}
\begin{aligned}
    e_q &= \frac{e_q^{graph}}{\|e_q^{graph}\|} + \frac{e_q^{search}}{\|e_q^{search}\|} + \frac{e_q^{joint}}{\|e_q^{joint}\|}, \;\;
    e_t = \frac{e_t^{graph}}{\|e_t^{graph}\|},
\end{aligned}
\end{equation}
\begin{equation}
\begin{aligned}
    s(q,t) = e_q^T \, e_t,
\end{aligned}
\end{equation}

\subsection{Learning Objective}


To optimize tool retrieval, we propose a sampled list-wise loss. For each query $q$, we construct a candidate tool set $\mathcal{S}$ of size $M$, containing all golden tools $t \in \mathcal{T}_q$ and $M - |\mathcal{T}_q|$ randomly sampled negatives from the global tool set. We define the ideal and predicted scoring distributions over tools and the sampled list-wise retrieval loss can be written as follows, where $\mathbb{I}(\cdot)$ is the indicator function:

\begin{equation}
    p_{q,t}={\mathbb{I}(t\in\mathcal{T}_q)}/{\sum\nolimits_{t^{\prime}\in\mathcal{S}}\mathbb{I}(t\in\mathcal{T}_q)}, \;\;t\in\mathcal{S},
\end{equation}

\begin{equation}
    \hat{p}_{q,t}={\exp(s(q,t))}/{\sum\nolimits_{t^{\prime}\in\mathcal{S}}\exp(s(q,t^{\prime}))},\;\;t\in\mathcal{S},
\end{equation}

\begin{equation}
    \mathcal{L}_{ret}=-\sum\nolimits_{t\in\mathcal{S}} p_{q,t}\log \hat{p}_{q,t}+(1-p_{q,t})\log(1-\hat{p}_{q,t}).
    \label{eq:total loss}
\end{equation}


Inspired by~\cite{qu2024colt}, we introduce an in-batch contrastive loss using the two query-centric graphs for regularization.
For each batch $\mathcal{B}$, we extract positive pairs: $(e_q^{QT}, e_q^{QS})$ for queries and $(e_s^{QT}, e_s^{QS})$ for scenes.
Here, $e_s^{QT}$ is computed via average pooling over the tools linked to the scene in the query-tool graph:
\(e_s^{QT} = \sum\nolimits_{t\in\mathcal{T}_q}e_t^{QT}/{|\mathcal{T}_q|}.\)
Based on the two pairs of positive representations, we adopt the InfoNCE~\cite{gutmann2010noise} loss as follows:
\begin{equation}
    \mathcal{L}_{con}= -\frac{1}{|\mathcal{B}|} \sum_{(q,s)\in\mathcal{B}}\left( \frac{e^{\operatorname{sim}(e_q^{QS},e_q^{QT})/\tau}}{\sum_{q^{\prime}}e^{\operatorname{sim}(e_q^{QS},e_{q^{\prime}}^{QT})/\tau}} + \frac{e^{\operatorname{sim}(e_s^{QS},e_s^{QT})/\tau}}{\sum_{s^{\prime}}e^{\operatorname{sim}(e_s^{QS},e_{s^{\prime}}^{QT})/\tau}} \right),
    \label{eq:contrastive loss}
\end{equation}
where $\tau$ is the temperature hyperparameter, and $\operatorname{sim}(\cdot,\cdot)$ is the dot product operation. Our MassTool can be optimized in an end-to-end manner with the learning objective:
\(\mathcal{L} = \mathcal{L}_{ret} + \lambda \mathcal{L}_{det} + \beta \mathcal{L}_{con},\)
where $\lambda$ and $\beta$ are the hyperparameters to balance the loss weights.

\section{Tool Usage Dataset Construction}

Existing datasets~\cite{qin2023toolllm,huang2023metatool,qu2024colt} fail to support the dual-step training of tool usage detection and tool retrieval. 
They only contain queries requiring tool invocations, ignoring the crucial step of determining whether the tool invocation is needed for the query. 
Hence, we construct the first tool usage detection dataset (dubbed \textbf{ToolDet}) based on the Chatbot Arena dataset~\cite{zheng2023judging} with the following systematic filtering and verification process:
\begin{itemize}[leftmargin=10pt]
    \item \textbf{Semantic Filtering}. 
    We first adopt Sentence-BERT~\cite{reimers2019sentence} to estimate the semantic similarity between each target query from the Chatbot Arena dataset and tool-dependent queries from other tool retrieval datasets (\eg, ToolLens~\cite{qu2024colt}). 
    We retain queries with the maximum similarity scores in the range between 0.4 and 0.6 for further verification.
    Queries in this range are likely to be tool-independent, while not being so obvious that they could be easily classified as tool-independent without detailed inspection. 
    This ensures the high-quality data for tool usage detection.
    \item \textbf{LLM \& Human Verification}. To ensure the filtered queries are genuinely tool-independent, we employ a proprietary large language model to analyze and verify that the queries could be resolved without invoking external tools.
    Moreover, we also engaged human annotators to manually review the remaining queries for the final round of verification.
    After this rigorous filtering process, we compiled 31,229 high-quality tool-independent queries to form the tool usage detection dataset.
\end{itemize}
ToolDet is dataset-agnostic and can complement existing tool retrieval datasets to enable the foundation for robust dual-step multi-task modeling. 
The dataset statistic of ToolDet is shown in Table~\ref{tab:datasets_Statistics}, and we have open-sourced the dataset to further benefit the broader research community. 
\section{Experiments}

\subsection{Experiment Setup}

\textbf{Datasets.} We conduct experiments on three widely adopted multi-tool retrieval datasets: ToolLens~\citep{qu2024colt}, ToolBenchG2~\citep{qin2023toolllm}, and ToolBenchG3~\citep{qin2023toolllm}. 
For MassTool, we use our ToolDet dataset as the complementary part for these datasets to enable the dual-step multi-task training of both tool usage detection and tool retrieval.
The detailed dataset statistics can be found in Appendix~\ref{app:dataset statistics}.

\textbf{Evaluation Metrics.} Following previous works~\citep{qu2024colt,zheng2024toolrerank,huang2023metatool}, we use the widely adopted Recall@K and NDCG@K as metrics to evaluate the performance of tool retrieval, with $K \in  \{3, 5\}$. 

\textbf{Baselines.} We adopt \textbf{ANCE}~\citep{xiong2020approximate}, \textbf{TAS-B}~\citep{hofstatter2021efficiently}, \textbf{coCondensor}~\citep{gao2021unsupervised} and \textbf{Contriever}~\citep{izacard2021unsupervised} as the backbone retrieval models for evaluation. 
Building upon these backbone models, we select four model-agnostic tool retrieval methods as our baselines: \textbf{QTA}~\citep{zhang2024data} ,\textbf{MMRR}~\citep{kachuee2024improving} , \textbf{APIRetriever}~\citep{qin2023toolllm} and \textbf{COLT}~\citep{qu2024colt}. A detailed description of
baselines can be referred to in Appendix~\ref{app: Baselines}.

\textbf{Implementation Details.} Due to page limitations, we give implementation details in Appendix~\ref{app:implementation}.


\begin{table}
\caption{Overall performance for tool retrieval.
The ``Raw'' framework means that the backbone is used directly for tool retrieval without finetuning.
The best result is given in bold, and the second-best value is underlined. 
The symbol $\ast$ indicates statistically significant improvement over the best baseline under t-test with $p$-value < 0.001. 
R and N stand for Recall and NDCG, respectively.
}
\label{tab:masstool_main}
\resizebox{\textwidth}{!}{
\renewcommand\arraystretch{1.07}
\begin{tabular}{c|c|cccc|cccc|cccc}
\toprule
\hline

\multirow{2}{*}{Backbone} & \multirow{2}{*}{Framework} & \multicolumn{4}{c|}{ToolLens} & \multicolumn{4}{c|}{ToolBenchG2} & \multicolumn{4}{c}{ToolBenchG3} \\  
\cline{3-14}
 & & R@3  & R@5 & N@3 & N@5 & R@3  & R@5 & N@3 & N@5 & R@3  & R@5 & N@3 & N@5\\ 
   \hline 
   
\multicolumn{1}{c|}{BM25} & - & 0.2158 & 0.2688 & 0.2319 & 0.2609 & 0.1706 & 0.2138 & 0.1783 & 0.1988 & 0.2933 &  0.3588 & 0.3220 & 0.3508 \\
\hline

\multicolumn{1}{c|}{\multirow{7}{*}{ANCE}} & Raw & 0.2083 & 0.2656 & 0.2145 & 0.2457 & 0.2083 & 0.2656 & 0.2145 & 0.2457 & 0.2155 &  0.2638 & 0.2344 & 0.2560 \\
\multicolumn{1}{c|}{\multirow{7}{*}{}} & QTA & 0.7718 & 0.9051 & 0.7892 & 0.8652 & 0.5545 & 0.6729 & 0.5822 & 0.6383 & 0.6408 & 0.7541 & 0.6855 & 0.7329 \\
\multicolumn{1}{c|}{\multirow{7}{*}{}} & MMRR & 0.7591 & 0.8992 & 0.7726 & 0.8524 & 0.5674 & 0.6839 & 0.5947 & 0.6524 & 0.6226 & 0.7422 & 0.6703 & 0.7217\\
\multicolumn{1}{c|}{\multirow{7}{*}{}} & APIRetriever & 0.8062 & 0.9417 & 0.8235 & 0.9015 & 0.5858 & 0.6720 & 0.5858 & 0.6375 & 0.6511 & 0.7663 & 0.6927 & 0.7414  \\
\multicolumn{1}{c|}{\multirow{7}{*}{}} & COLT & \underline{0.9215} & \underline{0.9778} & \underline{0.9278} & \underline{0.9610} & \underline{0.7076} & \underline{0.8059} & \underline{0.7076} & \underline{0.7798} & \underline{0.7337} & \underline{0.8397} & \underline{0.7795} & \underline{0.8214}  \\
\multicolumn{1}{c|}{\multirow{7}{*}{}} & \modelname~(Ours) & \textbf{0.9648$^*$} & \textbf{0.9847$^*$} & \textbf{0.9670$^*$} & \textbf{0.9785$^*$} & \textbf{0.7927$^*$} & \textbf{0.8678$^*$} & \textbf{0.8124$^*$} & \textbf{0.8429$^*$} & \textbf{0.7840$^*$} & \textbf{0.8662$^*$} & \textbf{0.8259$^*$} & \textbf{0.8540$^*$}  \\
\multicolumn{1}{c|}{\multirow{7}{*}{}} & Rel.Imprv. & 4.69\% & 0.71\% & 4.23\% & 1.82\% & 12.02\% & 7.68\% & 10.32\% & 8.09\% & 6.86\% & 3.16\% & 5.95\% & 3.97\% \\
   \hline  

\multicolumn{1}{c|}{\multirow{7}{*}{TAS-B}} & Raw & 0.1910 & 0.2371 & 0.1981 & 0.2233 & 0.1910 & 0.2371 & 0.1981 & 0.2233 & 0.2532 & 0.3115 & 0.2780 & 0.3036 \\
\multicolumn{1}{c|}{\multirow{7}{*}{}} & QTA & 0.7731 & 0.9031 & 0.7883 & 0.8623 & 0.5736 & 0.6872 & 0.6033 & 0.6561 & 0.6497 & 0.7637 & 0.6964 & 0.7443 \\
\multicolumn{1}{c|}{\multirow{7}{*}{}} & MMRR & 0.7607 & 0.8893 & 0.7785 & 0.8518 & 0.5786 & 0.6982 & 0.6061 & 0.6626 & 0.6419 & 0.7602 & 0.6844 & 0.7354\\
\multicolumn{1}{c|}{\multirow{7}{*}{}} & APIRetriever & 0.8126 & 0.9406 & 0.8254 & 0.8994 & 0.6278 & 0.6749 & 0.5896 & 0.6421 & 0.6604 & 0.7764 & 0.7041 & 0.7534 \\
\multicolumn{1}{c|}{\multirow{7}{*}{}} & COLT & \underline{0.9149} &  \underline{0.9691} &  \underline{0.9248} &  \underline{0.9563} &  \underline{0.7164} &  \underline{0.8112} &  \underline{0.7460} &  \underline{0.7874} &  \underline{0.7449} &  \underline{0.8458} &  \underline{0.7903} &  \underline{0.8295} \\
\multicolumn{1}{c|}{\multirow{7}{*}{}} & \modelname~(Ours) & \textbf{0.9523$^*$} & \textbf{0.9812$^*$} & \textbf{0.9577$^*$} & \textbf{0.9744$^*$} & \textbf{0.7958$^*$} & \textbf{0.8684$^*$} & \textbf{0.8164$^*$} & \textbf{0.8455$^*$} & \textbf{0.7923$^*$} & \textbf{0.8675$^*$} & \textbf{0.8338$^*$} & \textbf{0.8577$^*$} \\
\multicolumn{1}{c|}{\multirow{7}{*}{}} & Rel.Imprv. & 4.08\% & 1.25\% & 3.56\% & 1.89\% & 11.08\% & 7.05\% & 9.44\% & 7.38\% & 6.36\% & 2.57\% & 5.50\% & 3.40\% \\
   \hline  

\multicolumn{1}{c|}{\multirow{7}{*}{coCondensor}} & Raw & 0.1533 & 0.1937 & 0.1615 & 0.1832 & 0.1533 & 0.1937 & 0.1615 & 0.1832 & 0.2080 & 0.2524 & 0.2321 & 0.2515 \\
\multicolumn{1}{c|}{\multirow{7}{*}{}} & QTA & 0.7829 & 0.9139 & 0.7991 & 0.8738 & 0.5951 & 0.7163 & 0.6210 & 0.6778 & 0.6506 & 0.7718 & 0.6939 & 0.7454\\
\multicolumn{1}{c|}{\multirow{7}{*}{}} & MMRR & 0.7802 & 0.9072 & 0.7947 & 0.8677 & 0.5874 & 0.7121 & 0.6161 & 0.6749 & 0.6396 & 0.7573 & 0.6845 & 0.7345 \\
\multicolumn{1}{c|}{\multirow{7}{*}{}} & APIRetriever & 0.8237 & 0.9469 & 0.8390 & 0.9106 & 0.5770 & 0.6946 & 0.6080 & 0.6607 & 0.6697 & 0.7930 & 0.7120 & 0.7650 \\
\multicolumn{1}{c|}{\multirow{7}{*}{}} & COLT &  \underline{0.9265} &  \underline{0.9778} &  \underline{0.9316} &  \underline{0.9617} &  \underline{0.7391} &  \underline{0.8347} &  \underline{0.7675} &  \underline{0.8087} &  \underline{0.7548} &  \underline{0.8497} &  \underline{0.8000} &  \underline{0.8355} \\
\multicolumn{1}{c|}{\multirow{7}{*}{}} & \modelname~(Ours) & \textbf{0.9642$^*$} & \textbf{0.9847$^*$} & \textbf{0.9683$^*$} & \textbf{0.9802$^*$} & \textbf{0.8028$^*$} & \textbf{0.8823$^*$} & \textbf{0.8237$^*$} & \textbf{0.8562$^*$} & \textbf{0.7971$^*$} & \textbf{0.8742$^*$} & \textbf{0.8394$^*$} & \textbf{0.8645$^*$} \\
\multicolumn{1}{c|}{\multirow{7}{*}{}} & Rel.Imprv. & 4.07\% & 0.71\% & 3.94\% & 1.89\% & 8.62\% & 5.70\% & 7.32\% & 5.67\% & 5.60\% & 2.88\% & 4.93\% & 3.47\% \\
   \hline  
\multicolumn{1}{c|}{\multirow{7}{*}{Contriever}} & Raw & 0.2567 & 0.3115 & 0.2696 & 0.2995 & 0.2567 & 0.3115 & 0.2696 & 0.2995 & 0.3137 & 0.3860 & 0.3413 & 0.3737 \\
\multicolumn{1}{c|}{\multirow{7}{*}{}} & QTA & 0.7798 & 0.8988 & 0.7958 & 0.8631 & 0.6040 & 0.7253 & 0.6311 & 0.6880 & 0.6655 & 0.7790 & 0.7112 & 0.7583 \\
\multicolumn{1}{c|}{\multirow{7}{*}{}} & MMRR & 0.7454 & 0.8725 & 0.7712 & 0.8435 & 0.6109 & 0.7326 & 0.6398 & 0.6971 & 0.6618 & 0.7741 & 0.7077 & 0.7546 \\
\multicolumn{1}{c|}{\multirow{7}{*}{}} & APIRetriever & 0.8358 & 0.9517 & 0.8498 & 0.9169 & 0.5889 & 0.7075 & 0.6211 & 0.6742 & 0.6858 & 0.8005 & 0.7286 & 0.7769 \\
\multicolumn{1}{c|}{\multirow{7}{*}{}} & COLT &  \underline{0.9364} &  \underline{0.9775} &  \underline{0.9453} &  \underline{0.9691} &  \underline{0.7572} &  \underline{0.8503} &  \underline{0.7857} &  \underline{0.8254} &  \underline{0.7663} &  \underline{0.8550} &  \underline{0.8121} &  \underline{0.8418} \\
\multicolumn{1}{c|}{\multirow{7}{*}{}} & \modelname~(Ours) & \textbf{0.9616$^*$} & \textbf{0.9839$^*$} & \textbf{0.9660$^*$} & \textbf{0.9787$^*$} & \textbf{0.8219$^*$}& \textbf{0.8963$^*$} & \textbf{0.8432$^*$} & \textbf{0.8723$^*$} & \textbf{0.8163$^*$} & \textbf{0.8896$^*$} & \textbf{0.8585$^*$} & \textbf{0.8805$^*$} \\
\multicolumn{1}{c|}{\multirow{7}{*}{}} & Rel.Imprv. & 2.69\% & 0.65\% & 2.19\% & 0.99\% & 8.55\% & 5.41\% & 7.32\% & 5.68\% & 6.53\% & 4.05\% & 5.71\% & 4.60\% \\
   \hline  
   
   \bottomrule          
\end{tabular}
}
\end{table}


\subsection{Overall Performance}
We evaluate the performance of MassTool in comparison to different backbone models and tool retrieval baselines, and report the results in Table~\ref{tab:masstool_main}. 
We can obtain the following observations:
\begin{itemize}[leftmargin=8pt]
    \item All tool retrieval methods significantly outperform raw backbone models, confirming the importance of tool retrieval fine-tuning for better user intent modeling and tool document understanding. 
    \item COLT achieves the best performance among all the baseline methods. 
    Built upon semantic learning, COLT further leverages graph neural networks to integrate high-order collaborative knowledge. 
    This enhances the tool representations and results in better performance over semantic-only baselines.
    \item Our proposed MassTool outperforms all the baselines across different datasets and backbones, validating its robustness and generalizability. 
    MassTool enables deep user intent mining through the integration of QC-GCN, SUIM, and AdaKT.
    In this way, we deepen the insights for user intent mining and thereby unleash the superior performance of MassTool for tool retrieval.
    
\end{itemize}

\begin{figure}[t]
  \centering
  \includegraphics[width=0.88\textwidth]{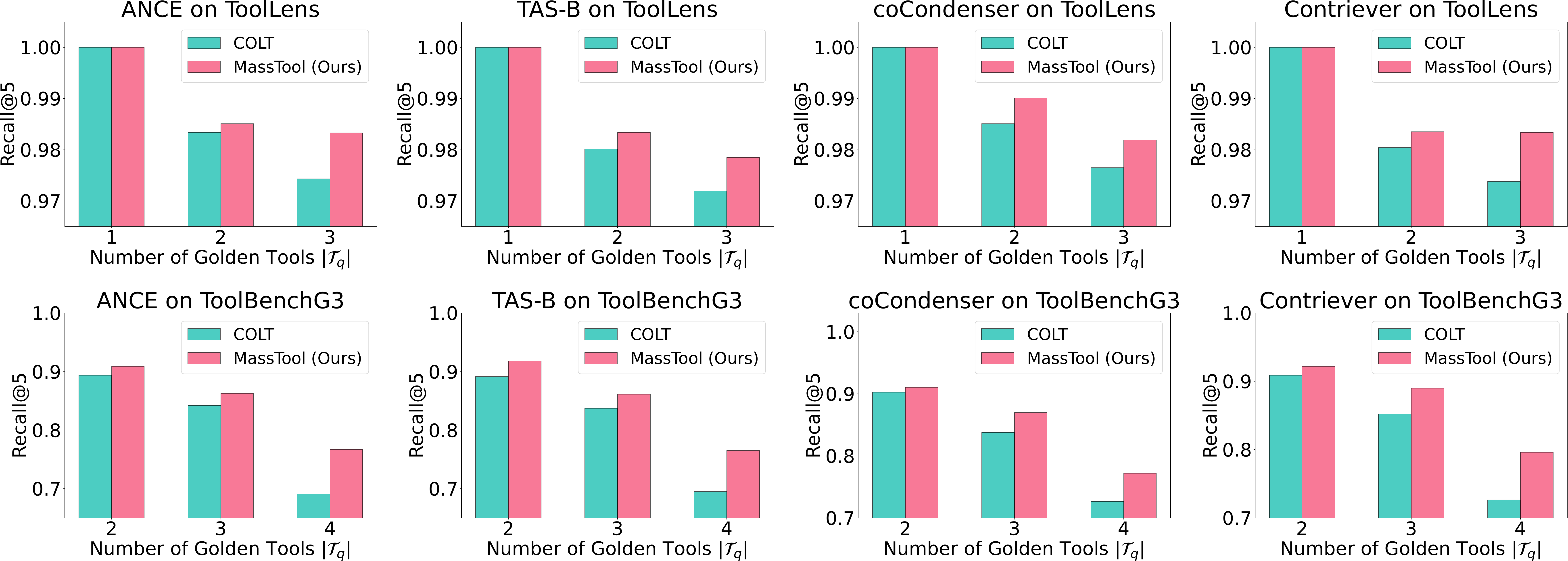}
  \caption{The performance of MassTool and COLT on different testing subsets w.r.t. the number of golden tools per query on ToolLens (top row) and ToolBenchG3 (bottom row) datasets.
  }
  \label{fig:differnet_N_G3}
\end{figure}

\subsection{Query Comprehension}

In this section, we further validate the capabilities of MassTool in comprehending complex queries and modeling diverse user intents.
The tool retrieval datasets we choose are all multi-tool datasets, where each input query would be associated with multiple tools.
Therefore, we choose ToolLens and ToolBenchG3 datasets, and divide their testing sets according to the number of golden tools per query, \ie, $|\mathcal{T}_q|$. 
Note that the greater the number of golden tools required by a query, the more complex the task it likely involves, which in turn reflects more intricate user intent for the query. 
We evaluate MassTool and the best baseline COLT on these testing subsets w.r.t. different numbers of golden tools per query, and report the results in Figure~\ref{fig:differnet_N_G3}.

Both the performances of MassTool and COLT decrease as the number of golden tools per query increases. 
This validates the hypothesis that a large number of golden tools generally indicates a more complex query to be resolved. 
Moreover, MassTool outperforms COLT on all the testing subsets. Moreover, the relative improvement becomes more pronounced as the number of golden tools increases (\ie, the complex level of query). 
This demonstrates that MassTool can effectively facilitate the understanding of complex queries and thereby improve the tool retrieval performance with enhanced query representations.


\subsection{Transfer Function}
\label{sec:transfer func exp}
In this section, we analyze the different choices of transfer function in the adaptive knowledge transfer module (AdaKT). 
In addition to the \textbf{element-wise gating} mechanism in our MassTool, given the inputs of detective knowledge $h$ and search-enhanced query representation $e_q^{search}$, we provide three alternative transfer functions: Attention, Concatenation and Addition. 
Further details of transfer functions are available in Appendix~\ref{app:transfer}.

We report the performance of different transfer functions in Table~\ref{tab:injection method}. 
The gating transfer function adopted by our MassTool generally achieves the best performance, except for a few cases. 
We argue that the design of gating mechanism is more aligned with the dual-step sequential decision-making nature for tool usage detection and tool retrieval. 
Based on the detective knowledge $h$ containing the potential information about whether the query should invoke tools or not, we can adaptively conduct element-wise gating for the search-based query representation $e_q^{search}$. 
In this way, we can selectively filter the unnecessary information and dynamically control the information flow for final prediction, leading to better knowledge fusion for multi-task modeling.

\begin{table}
\caption{The performance of MassTool w.r.t. different transfer functions in AdaKT.
}
\label{tab:injection method}
\resizebox{\textwidth}{!}{
\renewcommand\arraystretch{1.07}
\begin{tabular}{c|c|cccc|cccc|cccc}
\toprule
\hline

\multicolumn{1}{c|}{\multirow{2}{*}{Backbone}} & \multicolumn{1}{c|}{\multirow{2}{*}{Transfer Func.}}  & \multicolumn{4}{c|}{ToolLens} & \multicolumn{4}{c|}{ToolBenchG2} & \multicolumn{4}{c}{ToolBenchG3} \\ 
\cline{3-14}
\multicolumn{1}{c|}{} & \multicolumn{1}{c|}{} & R@3  & R@5 & N@3 & N@5 & R@3  & R@5 & N@3 & N@5 & R@3  & R@5 & N@3 & N@5\\ 
   \hline
   
\multicolumn{1}{c|}{\multirow{4}{*}{ANCE}} & \multicolumn{1}{c|}{Gating} & \textbf{0.9648} & \underline{0.9848} & \textbf{0.9670} & \underline{0.9785} & \textbf{0.7927} & \textbf{0.8678} & \underline{0.8124} & \textbf{0.8429} & \textbf{0.7840} & \textbf{0.8662} & \textbf{0.8259} & \textbf{0.8540}  \\
\multicolumn{1}{c|}{\multirow{4}{*}{}} & \multicolumn{1}{c|}{ Attention }& 0.9532 &  0.9827 & 0.9548 & 0.9721 & \underline{0.7926} &  \underline{0.8666} & \textbf{0.8134} & \underline{0.8428} & \underline{0.7813}  & \underline{0.8661} & \underline{0.8223} & \underline{0.8521} \\
\multicolumn{1}{c|}{\multirow{4}{*}{}} & \multicolumn{1}{c|}{ Concatenation } & 0.9622 & \textbf{0.9855} & \underline{0.9656} & \textbf{0.9791} & 0.7765 & 0.8524 & 0.7982 & 0.8293 & 0.7757 & 0.8617 & 0.8165 & 0.8471 \\
\multicolumn{1}{c|}{\multirow{4}{*}{}} & \multicolumn{1}{c|}{ Addition } & 0.9600 & \underline{0.9848} & 0.9623 & 0.9766 & 0.7540 & 0.8505 & 0.7748 & 0.8261 & 0.7769 & 0.8629 & 0.8181 & 0.8485 \\
   \hline  

\multicolumn{1}{c|}{\multirow{4}{*}{TAS-B}} & \multicolumn{1}{c|}{Gating} & \textbf{0.9523} & \textbf{0.9812} & \textbf{0.9577} & \textbf{0.9744} & \textbf{0.7958} & \textbf{0.8684} & \textbf{0.8164} & \textbf{0.8455} & \textbf{0.7923} & \underline{0.8675} & \textbf{0.8338} & \textbf{0.8577} \\
\multicolumn{1}{c|}{\multirow{4}{*}{}} & \multicolumn{1}{c|}{ Attention }& 0.9513 & 0.9794 & 0.9557 & 0.9720 & \underline{0.7940} & \underline{0.8667} & \underline{0.8114} & \underline{0.8432} & \underline{0.7909} & \textbf{0.8705} & \underline{0.8316} & \underline{0.8576} \\
\multicolumn{1}{c|}{\multirow{4}{*}{}} & \multicolumn{1}{c|}{ Concatenation } & \underline{0.9520} & 0.9808 & \underline{0.9566} & \underline{0.9740} & 0.7782 & 0.8516 & 0.7979 & 0.8278 & 0.7836 & 0.8663 & 0.8255 & 0.8542 \\
\multicolumn{1}{c|}{\multirow{4}{*}{}} & \multicolumn{1}{c|}{ Addition } & 0.9492 & \underline{0.9810} & 0.9523 & 0.9709 & 0.7775 & 0.8446 & 0.7994 & 0.8256 & 0.7832 & 0.8649 & 0.8271 & 0.8548 \\
   \hline  
\multicolumn{1}{c|}{\multirow{4}{*}{coCondensor}} & \multicolumn{1}{c|}{Gating} & \textbf{0.9642} & \underline{0.9847} & \underline{0.9683} & \textbf{0.9802} & \textbf{0.8028} & \textbf{0.8823} & \textbf{0.8237} & \textbf{0.8562} & \textbf{0.7971} & \underline{0.8742} & \textbf{0.8394} & \textbf{0.8645} \\
\multicolumn{1}{c|}{\multirow{4}{*}{}} & \multicolumn{1}{c|}{ Attention }& 0.9561 & 0.9832 & 0.9605 & 0.9762 & \underline{0.8004} & \underline{0.8798} & \underline{0.8203} & \underline{0.8531} & \underline{0.7964} & \textbf{0.8762} & \underline{0.8380} & \textbf{0.8645} \\
\multicolumn{1}{c|}{\multirow{4}{*}{}} & \multicolumn{1}{c|}{ Concatenation } & \underline{0.9641} & 0.9835 & \textbf{0.9687} & \underline{0.9795} &  0.7919 &  0.8781 &  0.8143 &  0.8507 &  0.7871 &  0.8662 &  0.8084 &  0.8410 \\
\multicolumn{1}{c|}{\multirow{4}{*}{}} & \multicolumn{1}{c|}{ Addition } & 0.9586 & \textbf{0.9857} & 0.9626 & 0.9784 & 0.7852 & 0.8622 & 0.8075 & 0.8390 & 0.7887 & 0.8729 & 0.8306 & 0.8596 \\
   \hline  

\multicolumn{1}{c|}{\multirow{4}{*}{Contriever}} & \multicolumn{1}{c|}{Gating} & \textbf{0.9616} &  \textbf{0.9839} &  \textbf{0.9660} &  \textbf{0.9787} & \textbf{0.8219}& \textbf{0.8963} & \textbf{0.8432} & \textbf{0.8723} & \textbf{0.8163} & \textbf{0.8896} & \textbf{0.8585} & \textbf{0.8805} \\
\multicolumn{1}{c|}{\multirow{4}{*}{}} & \multicolumn{1}{c|}{ Attention }& 0.9567 & \underline{0.9826} & \underline{0.9618} & \underline{0.9767} & \underline{0.8089} & \underline{0.8886} & \underline{0.8282} & \underline{0.8607} & \underline{0.8162} & \underline{0.8895} & \underline{0.8565} & \underline{0.8788} \\
\multicolumn{1}{c|}{\multirow{4}{*}{}} & \multicolumn{1}{c|}{ Concatenation} & \underline{0.9592} & 0.9819 & 0.9612 & 0.9745 &  0.7861 &  0.8569 &  0.8077 &  0.8358 &  0.7887 &  0.8721 &  0.8306 &  0.8592 \\
\multicolumn{1}{c|}{\multirow{4}{*}{}} & \multicolumn{1}{c|}{ Addition} & 0.9514 & 0.9811 & 0.9541 & 0.9712 & 0.7853 & 0.8735 & 0.8111 & 0.8484 & 0.8016 & 0.8775 & 0.8429 & 0.8666 \\

   \hline  
   \bottomrule          
\end{tabular}

}
\end{table}

\subsection{In-Depth Analysis}

We analyze the impact of different components in MassTool, \ie, ablation study. We conduct experiments with four backbone models on ToolLens and ToolBenchG3 datasets. We conduct the ablation study by evaluating the performance of the following model variants: MassTool w/o AdaKT, MassTool w/o SUIM, MassTool w/o DF (Dynamic Filtering), and MassTool w/o CL (Contrastive Learning). 
Further details of model variants are available in Appendix~\ref{app:ablation}.
We report results in Figure~\ref{fig:ablation} and draw the following observations:
\begin{itemize}[leftmargin=10pt]
    \item MassTool w/o AdaKT suffers from a considerable performance decrease, demonstrating the importance of casting the dual-step multi-task modeling of both tool usage detection and tool retrieval tasks. The information gained from the previous detective stage can establish a more comprehensive view of user queries and intents, thereby leading to better retrieval performance.
    \item MassTool w/o SUIM and w/o DF generally meet significant performance drops, validating the effectiveness of our search-based techniques to model the nuanced user intents. 
    \item The performance of MassTool w/o CL also declines to some extent. This shows the importance of regularization to better learn the query-tool-scene representations with multiple views, preventing the neural tool retriever from overfitting. 
\end{itemize}
\begin{figure}[t]
  \centering
  \includegraphics[width=0.83\textwidth]{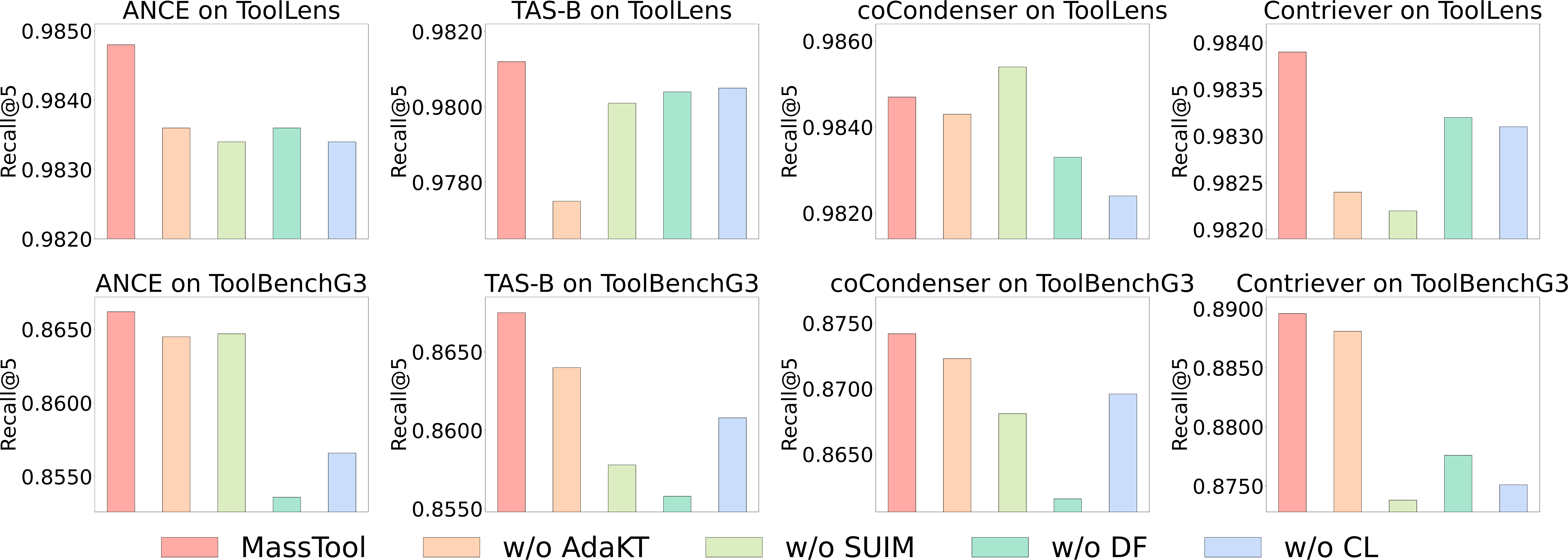}
  \caption{The performance of different MassTool variants on ToolLens (top row) and ToolBenchG3 (bottom row) datasets.
  }
  \label{fig:ablation}
\end{figure}

Due to the page limitation, we provide more analysis experiments in Appendix~\ref{app:addtional exp}, covering the hyperparameter studies and visualization of query representations, respectively.

\section{Conclusion and Limitation}
\label{sec: conclusion}
In this paper, we focus on tool retrieval for large language models (LLMs). 
We identify key challenges of query representation optimization, particularly in capturing nuanced user intent and addressing the dual-step sequential decision-making nature of tool retrieval. 
To this end, we propose a multi-task search-based tool retrieval framework (\ie, MassTool), featuring a two-tower architecture to enable the dual-step training of both tool usage detection and tool retrieval tasks. 
Moreover, we release the first tool usage detection dataset (\ie, ToolDet), which can be integrated with existing tool retrieval datasets for enhancements. 
Extensive experiments demonstrate the effectiveness of MassTool in enhancing query representations and improving tool retrieval performance. 
While effective, MassTool does not yet support continual tool expansion and does not interact with tool-augmented LLMs.
As for future work, we will explore the synergy between tool-augmented LLMs and external tools with the tool retriever as the bridge.

\medskip
{
\bibliographystyle{rusnat}
\bibliography{acmart}

\begin{thebibliography}{56}
\providecommand{\natexlab}[1]{#1}
\providecommand{\EM}{\em}
\providecommand{\RNtxt}{\relax}
\RNtxt{}

\bibitem[Agarap(2018)A.~Agarap]{agarap2018deep}
{\EM Agarap AF}.
\newblock Deep learning using rectified linear units (relu) \allowbreak\newblock// arXiv preprint arXiv:1803.08375. 2018.

\bibitem[Fore et~al.(2024)M.~Fore, S.~Singh, D.~Stamoulis]{fore2024geckopt}
{\EM Fore Michael, Singh Simranjit, Stamoulis Dimitrios}.
\newblock GeckOpt: LLM System Efficiency via Intent-Based Tool Selection \allowbreak\newblock// Proceedings of the Great Lakes Symposium on VLSI 2024. 2024.  353--354.

\bibitem[Gao, Zhang(2024)H.~Gao, Y.~Zhang]{gao2024ptr}
{\EM Gao Hang, Zhang Yongfeng}.
\newblock PTR: Precision-Driven Tool Recommendation for Large Language Models \allowbreak\newblock// arXiv preprint arXiv:2411.09613. 2024.

\bibitem[Gao, Callan(2021)L.~Gao, J.~Callan]{gao2021unsupervised}
{\EM Gao Luyu, Callan Jamie}.
\newblock Unsupervised corpus aware language model pre-training for dense passage retrieval \allowbreak\newblock// arXiv preprint arXiv:2108.05540. 2021.

\bibitem[Gao et~al.(2024)S.~Gao, Z.~Shi, M.~Zhu, B.~Fang, X.~Xin, P.~Ren, Z.~Chen, J.~Ma, Z.~Ren]{gao2024confucius}
{\EM Gao Shen, Shi Zhengliang, Zhu Minghang, Fang Bowen, Xin Xin, Ren Pengjie, Chen Zhumin, Ma~Jun, Ren Zhaochun}.
\newblock Confucius: Iterative tool learning from introspection feedback by easy-to-difficult curriculum \allowbreak\newblock// Proceedings of the AAAI Conference on Artificial Intelligence.  38, 16. 2024.  18030--18038.

\bibitem[Gutmann, Hyv{\"a}rinen(2010)M.~Gutmann, A.~Hyv{\"a}rinen]{gutmann2010noise}
{\EM Gutmann Michael, Hyv{\"a}rinen Aapo}.
\newblock Noise-contrastive estimation: A new estimation principle for unnormalized statistical models \allowbreak\newblock// Proceedings of the thirteenth international conference on artificial intelligence and statistics. 2010.  297--304.

\bibitem[He et~al.(2020)X.~He, K.~Deng, X.~Wang, Y.~Li, Y.~Zhang, M.~Wang]{he2020lightgcn}
{\EM He~Xiangnan, Deng Kuan, Wang Xiang, Li~Yan, Zhang Yongdong, Wang Meng}.
\newblock Lightgcn: Simplifying and powering graph convolution network for recommendation \allowbreak\newblock// Proceedings of the 43rd International ACM SIGIR conference on research and development in Information Retrieval. 2020.  639--648.

\bibitem[Hofst{\"a}tter et~al.(2021)S.~Hofst{\"a}tter, S.-C. Lin, J.-H. Yang, J.~Lin, A.~Hanbury]{hofstatter2021efficiently}
{\EM Hofst{\"a}tter Sebastian, Lin Sheng-Chieh, Yang Jheng-Hong, Lin Jimmy, Hanbury Allan}.
\newblock Efficiently teaching an effective dense retriever with balanced topic aware sampling \allowbreak\newblock// Proceedings of the 44th International ACM SIGIR Conference on Research and Development in Information Retrieval. 2021.  113--122.

\bibitem[Huang et~al.(2024{\natexlab{a}})J.~Huang, J.~Chen, J.~Lin, J.~Qin, Z.~Feng, W.~Zhang, Y.~Yu]{huang2024comprehensive}
{\EM Huang Junjie, Chen Jizheng, Lin Jianghao, Qin Jiarui, Feng Ziming, Zhang Weinan, Yu~Yong}.
\newblock A Comprehensive Survey on Retrieval Methods in Recommender Systems \allowbreak\newblock// arXiv preprint arXiv:2407.21022. 2024{\natexlab{a}}.

\bibitem[Huang et~al.(2024{\natexlab{b}})S.~Huang, W.~Zhong, J.~Lu, Q.~Zhu, J.~Gao, W.~Liu, Y.~Hou, X.~Zeng, Y.~Wang, L.~Shang, et~al.]{huang2024planning}
{\EM Huang Shijue, Zhong Wanjun, Lu~Jianqiao, Zhu Qi, Gao Jiahui, Liu Weiwen, Hou Yutai, Zeng Xingshan, Wang Yasheng, Shang Lifeng, others }.
\newblock Planning, Creation, Usage: Benchmarking LLMs for Comprehensive Tool Utilization in Real-World Complex Scenarios \allowbreak\newblock// arXiv preprint arXiv:2401.17167. 2024{\natexlab{b}}.

\bibitem[Huang et~al.(2023)Y.~Huang, J.~Shi, Y.~Li, C.~Fan, S.~Wu, Q.~Zhang, Y.~Liu, P.~Zhou, Y.~Wan, N.~Z. Gong, et~al.]{huang2023metatool}
{\EM Huang Yue, Shi Jiawen, Li~Yuan, Fan Chenrui, Wu~Siyuan, Zhang Qihui, Liu Yixin, Zhou Pan, Wan Yao, Gong Neil~Zhenqiang, others }.
\newblock Metatool benchmark for large language models: Deciding whether to use tools and which to use \allowbreak\newblock// arXiv preprint arXiv:2310.03128. 2023.

\bibitem[Izacard et~al.(2021)G.~Izacard, M.~Caron, L.~Hosseini, S.~Riedel, P.~Bojanowski, A.~Joulin, E.~Grave]{izacard2021unsupervised}
{\EM Izacard Gautier, Caron Mathilde, Hosseini Lucas, Riedel Sebastian, Bojanowski Piotr, Joulin Armand, Grave Edouard}.
\newblock Unsupervised dense information retrieval with contrastive learning \allowbreak\newblock// arXiv preprint arXiv:2112.09118. 2021.

\bibitem[Joko et~al.(2024)H.~Joko, S.~Chatterjee, A.~Ramsay, A.~P. De~Vries, J.~Dalton, F.~Hasibi]{joko2024doing}
{\EM Joko Hideaki, Chatterjee Shubham, Ramsay Andrew, De~Vries Arjen~P, Dalton Jeff, Hasibi Faegheh}.
\newblock Doing personal laps: Llm-augmented dialogue construction for personalized multi-session conversational search \allowbreak\newblock// Proceedings of the 47th International ACM SIGIR Conference on Research and Development in Information Retrieval. 2024.  796--806.

\bibitem[Joshi et~al.(2024)A.~Joshi, R.~Dabre, D.~Kanojia, Z.~Li, H.~Zhan, G.~Haffari, D.~Dippold]{joshi2024natural}
{\EM Joshi Aditya, Dabre Raj, Kanojia Diptesh, Li~Zhuang, Zhan Haolan, Haffari Gholamreza, Dippold Doris}.
\newblock Natural language processing for dialects of a language: A survey \allowbreak\newblock// arXiv preprint arXiv:2401.05632. 2024.

\bibitem[Kachuee et~al.(2024)M.~Kachuee, S.~Ahuja, V.~Kumar, P.~Xu, X.~Liu]{kachuee2024improving}
{\EM Kachuee Mohammad, Ahuja Sarthak, Kumar Vaibhav, Xu~Puyang, Liu Xiaohu}.
\newblock Improving Tool Retrieval by Leveraging Large Language Models for Query Generation \allowbreak\newblock// arXiv preprint arXiv:2412.03573. 2024.

\bibitem[Li et~al.(2023{\natexlab{a}})M.~Li, Y.~Zhao, B.~Yu, F.~Song, H.~Li, H.~Yu, Z.~Li, F.~Huang, Y.~Li]{li2023api}
{\EM Li~Minghao, Zhao Yingxiu, Yu~Bowen, Song Feifan, Li~Hangyu, Yu~Haiyang, Li~Zhoujun, Huang Fei, Li~Yongbin}.
\newblock Api-bank: A comprehensive benchmark for tool-augmented llms \allowbreak\newblock// arXiv preprint arXiv:2304.08244. 2023{\natexlab{a}}.

\bibitem[Li et~al.(2023{\natexlab{b}})X.~Li, L.~Su, P.~Jia, X.~Zhao, S.~Cheng, J.~Wang, D.~Yin]{li2023agent4ranking}
{\EM Li~Xiaopeng, Su~Lixin, Jia Pengyue, Zhao Xiangyu, Cheng Suqi, Wang Junfeng, Yin Dawei}.
\newblock Agent4ranking: Semantic robust ranking via personalized query rewriting using multi-agent llm \allowbreak\newblock// arXiv preprint arXiv:2312.15450. 2023{\natexlab{b}}.

\bibitem[Lin et~al.(2025)J.~Lin, X.~Dai, Y.~Xi, W.~Liu, B.~Chen, H.~Zhang, Y.~Liu, C.~Wu, X.~Li, C.~Zhu, et~al.]{lin2025can}
{\EM Lin Jianghao, Dai Xinyi, Xi~Yunjia, Liu Weiwen, Chen Bo, Zhang Hao, Liu Yong, Wu~Chuhan, Li~Xiangyang, Zhu Chenxu, others }.
\newblock How can recommender systems benefit from large language models: A survey \allowbreak\newblock// ACM Transactions on Information Systems. 2025. 43, 2. 1--47.

\bibitem[Lin et~al.(2021)J.~Lin, W.~Liu, X.~Dai, W.~Zhang, S.~Li, R.~Tang, X.~He, J.~Hao, Y.~Yu]{lin2021graph}
{\EM Lin Jianghao, Liu Weiwen, Dai Xinyi, Zhang Weinan, Li~Shuai, Tang Ruiming, He~Xiuqiang, Hao Jianye, Yu~Yong}.
\newblock A Graph-Enhanced Click Model for Web Search \allowbreak\newblock// Proceedings of the 44th International ACM SIGIR Conference on Research and Development in Information Retrieval. 2021.  1259--1268.

\bibitem[Lin et~al.(2024)Q.~Lin, M.~Wen, Q.~Peng, G.~Nie, J.~Liao, J.~Wang, X.~Mo, J.~Zhou, C.~Cheng, Y.~Zhao, et~al.]{lin2024hammer}
{\EM Lin Qiqiang, Wen Muning, Peng Qiuying, Nie Guanyu, Liao Junwei, Wang Jun, Mo~Xiaoyun, Zhou Jiamu, Cheng Cheng, Zhao Yin, others }.
\newblock Hammer: Robust Function-Calling for On-Device Language Models via Function Masking \allowbreak\newblock// arXiv preprint arXiv:2410.04587. 2024.

\bibitem[Liu et~al.(2024{\natexlab{a}})W.~Liu, X.~Huang, X.~Zeng, X.~Hao, S.~Yu, D.~Li, S.~Wang, W.~Gan, Z.~Liu, Y.~Yu, et~al.]{liu2024toolace}
{\EM Liu Weiwen, Huang Xu, Zeng Xingshan, Hao Xinlong, Yu~Shuai, Li~Dexun, Wang Shuai, Gan Weinan, Liu Zhengying, Yu~Yuanqing, others }.
\newblock ToolACE: Winning the Points of LLM Function Calling \allowbreak\newblock// arXiv preprint arXiv:2409.00920. 2024{\natexlab{a}}.

\bibitem[Liu et~al.(2024{\natexlab{b}})W.~Liu, Y.~Zhu, Z.~Dou]{liu2024demorank}
{\EM Liu Wenhan, Zhu Yutao, Dou Zhicheng}.
\newblock DemoRank: Selecting Effective Demonstrations for Large Language Models in Ranking Task \allowbreak\newblock// arXiv preprint arXiv:2406.16332. 2024{\natexlab{b}}.

\bibitem[Liu et~al.(2024{\natexlab{c}})Y.~Liu, Y.~Yuan, C.~Wang, J.~Han, Y.~Ma, L.~Zhang, N.~Zheng, H.~Xu]{liu2024summary}
{\EM Liu Yulong, Yuan Yunlong, Wang Chunwei, Han Jianhua, Ma~Yongqiang, Zhang Li, Zheng Nanning, Xu~Hang}.
\newblock From summary to action: Enhancing large language models for complex tasks with open world apis \allowbreak\newblock// arXiv preprint arXiv:2402.18157. 2024{\natexlab{c}}.

\bibitem[Liu et~al.(2024{\natexlab{d}})Z.~Liu, T.~Hoang, J.~Zhang, M.~Zhu, T.~Lan, S.~Kokane, J.~Tan, W.~Yao, Z.~Liu, Y.~Feng, et~al.]{liu2024apigen}
{\EM Liu Zuxin, Hoang Thai, Zhang Jianguo, Zhu Ming, Lan Tian, Kokane Shirley, Tan Juntao, Yao Weiran, Liu Zhiwei, Feng Yihao, others }.
\newblock Apigen: Automated pipeline for generating verifiable and diverse function-calling datasets \allowbreak\newblock// arXiv preprint arXiv:2406.18518. 2024{\natexlab{d}}.

\bibitem[Luo et~al.(2024)L.~Luo, Y.~Liu, R.~Liu, S.~Phatale, H.~Lara, Y.~Li, L.~Shu, Y.~Zhu, L.~Meng, J.~Sun, et~al.]{luo2024improve}
{\EM Luo Liangchen, Liu Yinxiao, Liu Rosanne, Phatale Samrat, Lara Harsh, Li~Yunxuan, Shu Lei, Zhu Yun, Meng Lei, Sun Jiao, others }.
\newblock Improve Mathematical Reasoning in Language Models by Automated Process Supervision \allowbreak\newblock// arXiv preprint arXiv:2406.06592. 2024.

\bibitem[Maaten Van~der, Hinton(2008)L.~Van~der Maaten, G.~Hinton]{van2008visualizing}
{\EM Maaten Laurens Van~der, Hinton Geoffrey}.
\newblock Visualizing data using t-SNE. \allowbreak\newblock// Journal of machine learning research. 2008. 9, 11.

\bibitem[Moon et~al.(2024)S.~Moon, S.~Jha, L.~E. Erdogan, S.~Kim, W.~Lim, K.~Keutzer, A.~Gholami]{moon2024efficient}
{\EM Moon Suhong, Jha Siddharth, Erdogan Lutfi~Eren, Kim Sehoon, Lim Woosang, Keutzer Kurt, Gholami Amir}.
\newblock Efficient and Scalable Estimation of Tool Representations in Vector Space \allowbreak\newblock// arXiv preprint arXiv:2409.02141. 2024.

\bibitem[Mu et~al.(2024)F.~Mu, Y.~Jiang, L.~Zhang, C.~Liu, W.~Li, P.~Xie, F.~Huang]{mu2024adaptive}
{\EM Mu~Feiteng, Jiang Yong, Zhang Liwen, Liu Chu, Li~Wenjie, Xie Pengjun, Huang Fei}.
\newblock Adaptive Selection for Homogeneous Tools: An Instantiation in the RAG Scenario \allowbreak\newblock// arXiv preprint arXiv:2406.12429. 2024.

\bibitem[Parisi et~al.(2022)A.~Parisi, Y.~Zhao, N.~Fiedel]{parisi2022talm}
{\EM Parisi Aaron, Zhao Yao, Fiedel Noah}.
\newblock Talm: Tool augmented language models \allowbreak\newblock// arXiv preprint arXiv:2205.12255. 2022.

\bibitem[Patil et~al.(2023)S.~G. Patil, T.~Zhang, X.~Wang, J.~E. Gonzalez]{patil2023gorilla}
{\EM Patil Shishir~G, Zhang Tianjun, Wang Xin, Gonzalez Joseph~E}.
\newblock Gorilla: Large language model connected with massive apis \allowbreak\newblock// arXiv preprint arXiv:2305.15334. 2023.

\bibitem[Peng et~al.(2024)W.~Peng, G.~Li, Y.~Jiang, Z.~Wang, D.~Ou, X.~Zeng, D.~Xu, T.~Xu, E.~Chen]{peng2024large}
{\EM Peng Wenjun, Li~Guiyang, Jiang Yue, Wang Zilong, Ou~Dan, Zeng Xiaoyi, Xu~Derong, Xu~Tong, Chen Enhong}.
\newblock Large language model based long-tail query rewriting in taobao search \allowbreak\newblock// Companion Proceedings of the ACM on Web Conference 2024. 2024.  20--28.

\bibitem[Qin et~al.(2023)Y.~Qin, S.~Liang, Y.~Ye, K.~Zhu, L.~Yan, Y.~Lu, Y.~Lin, X.~Cong, X.~Tang, B.~Qian, S.~Zhao, R.~Tian, R.~Xie, J.~Zhou, M.~Gerstein, D.~Li, Z.~Liu, M.~Sun]{qin2023toolllm}
{\EM Qin Yujia, Liang Shihao, Ye~Yining, Zhu Kunlun, Yan Lan, Lu~Yaxi, Lin Yankai, Cong Xin, Tang Xiangru, Qian Bill, Zhao Sihan, Tian Runchu, Xie Ruobing, Zhou Jie, Gerstein Mark, Li~Dahai, Liu Zhiyuan, Sun Maosong}.
\newblock ToolLLM: Facilitating Large Language Models to Master 16000+ Real-world APIs. 2023.

\bibitem[Qu et~al.(2024{\natexlab{a}})C.~Qu, S.~Dai, X.~Wei, H.~Cai, S.~Wang, D.~Yin, J.~Xu, J.-R. Wen]{qu2024tool}
{\EM Qu~Changle, Dai Sunhao, Wei Xiaochi, Cai Hengyi, Wang Shuaiqiang, Yin Dawei, Xu~Jun, Wen Ji-Rong}.
\newblock Tool Learning with Large Language Models: A Survey \allowbreak\newblock// arXiv preprint arXiv:2405.17935. 2024{\natexlab{a}}.

\bibitem[Qu et~al.(2024{\natexlab{b}})C.~Qu, S.~Dai, X.~Wei, H.~Cai, S.~Wang, D.~Yin, J.~Xu, J.-R. Wen]{qu2024colt}
{\EM Qu~Changle, Dai Sunhao, Wei Xiaochi, Cai Hengyi, Wang Shuaiqiang, Yin Dawei, Xu~Jun, Wen Ji-Rong}.
\newblock Towards completeness-oriented tool retrieval for large language models \allowbreak\newblock// Proceedings of the 33rd ACM International Conference on Information and Knowledge Management. 2024{\natexlab{b}}.  1930--1940.

\bibitem[Reimers, Gurevych(2019)N.~Reimers, I.~Gurevych]{reimers2019sentence}
{\EM Reimers Nils, Gurevych Iryna}.
\newblock Sentence-bert: Sentence embeddings using siamese bert-networks \allowbreak\newblock// arXiv preprint arXiv:1908.10084. 2019.

\bibitem[Robertson et~al.(2009)S.~Robertson, H.~Zaragoza, et~al.]{robertson2009probabilistic}
{\EM Robertson Stephen, Zaragoza Hugo, others }.
\newblock The probabilistic relevance framework: BM25 and beyond \allowbreak\newblock// Foundations and Trends{\textregistered} in Information Retrieval. 2009. 3, 4. 333--389.

\bibitem[Satpute et~al.(2024)A.~Satpute, N.~Gie{\ss}ing, A.~Greiner-Petter, M.~Schubotz, O.~Teschke, A.~Aizawa, B.~Gipp]{satpute2024can}
{\EM Satpute Ankit, Gie{\ss}ing Noah, Greiner-Petter Andr{\'e}, Schubotz Moritz, Teschke Olaf, Aizawa Akiko, Gipp Bela}.
\newblock Can llms master math? investigating large language models on math stack exchange \allowbreak\newblock// Proceedings of the 47th International ACM SIGIR Conference on Research and Development in Information Retrieval. 2024.  2316--2320.

\bibitem[Schick et~al.(2024)T.~Schick, J.~Dwivedi-Yu, R.~Dess{\`\i}, R.~Raileanu, M.~Lomeli, E.~Hambro, L.~Zettlemoyer, N.~Cancedda, T.~Scialom]{schick2024toolformer}
{\EM Schick Timo, Dwivedi-Yu Jane, Dess{\`\i} Roberto, Raileanu Roberta, Lomeli Maria, Hambro Eric, Zettlemoyer Luke, Cancedda Nicola, Scialom Thomas}.
\newblock Toolformer: Language models can teach themselves to use tools \allowbreak\newblock// Advances in Neural Information Processing Systems. 2024. 36.

\bibitem[Shen et~al.(2024)Y.~Shen, K.~Song, X.~Tan, D.~Li, W.~Lu, Y.~Zhuang]{shen2024hugginggpt}
{\EM Shen Yongliang, Song Kaitao, Tan Xu, Li~Dongsheng, Lu~Weiming, Zhuang Yueting}.
\newblock Hugginggpt: Solving ai tasks with chatgpt and its friends in hugging face \allowbreak\newblock// Advances in Neural Information Processing Systems. 2024. 36.

\bibitem[Siro et~al.(2024)C.~Siro, M.~Aliannejadi, M.~de~Rijke]{siro2024rethinking}
{\EM Siro Clemencia, Aliannejadi Mohammad, Rijke Maarten de}.
\newblock Rethinking the evaluation of dialogue systems: Effects of user feedback on crowdworkers and LLMs \allowbreak\newblock// Proceedings of the 47th International ACM SIGIR Conference on Research and Development in Information Retrieval. 2024.  1952--1962.

\bibitem[Sparck~Jones(1972)K.~Sparck~Jones]{sparck1972statistical}
{\EM Sparck~Jones Karen}.
\newblock A statistical interpretation of term specificity and its application in retrieval \allowbreak\newblock// Journal of documentation. 1972. 28, 1. 11--21.

\bibitem[Tang et~al.(2023)Q.~Tang, Z.~Deng, H.~Lin, X.~Han, Q.~Liang, B.~Cao, L.~Sun]{tang2023toolalpaca}
{\EM Tang Qiaoyu, Deng Ziliang, Lin Hongyu, Han Xianpei, Liang Qiao, Cao Boxi, Sun Le}.
\newblock Toolalpaca: Generalized tool learning for language models with 3000 simulated cases \allowbreak\newblock// arXiv preprint arXiv:2306.05301. 2023.

\bibitem[Thakur et~al.(2021)N.~Thakur, N.~Reimers, A.~Rücklé, A.~Srivastava, I.~Gurevych]{thakur2021beirheterogenousbenchmarkzeroshot}
{\EM Thakur Nandan, Reimers Nils, Rücklé Andreas, Srivastava Abhishek, Gurevych Iryna}.
\newblock BEIR: A Heterogenous Benchmark for Zero-shot Evaluation of Information Retrieval Models. 2021.

\bibitem[Vaswani et~al.(2017)A.~Vaswani, N.~Shazeer, N.~Parmar, J.~Uszkoreit, L.~Jones, A.~N. Gomez, {\L}.~Kaiser, I.~Polosukhin]{vaswani2017attention}
{\EM Vaswani Ashish, Shazeer Noam, Parmar Niki, Uszkoreit Jakob, Jones Llion, Gomez Aidan~N, Kaiser {\L}ukasz, Polosukhin Illia}.
\newblock Attention is all you need \allowbreak\newblock// Advances in neural information processing systems. 2017.  5998--6008.

\bibitem[Wang et~al.(2024)C.~Wang, W.~Luo, Q.~Chen, H.~Mai, J.~Guo, S.~Dong, Z.~Li, L.~Ma, S.~Gao, et~al.]{wang2024tool}
{\EM Wang Chenyu, Luo Weixin, Chen Qianyu, Mai Haonan, Guo Jindi, Dong Sixun, Li~Zhengxin, Ma~Lin, Gao Shenghua, others }.
\newblock Tool-LMM: A Large Multi-Modal Model for Tool Agent Learning \allowbreak\newblock// arXiv preprint arXiv:2401.10727. 2024.

\bibitem[Wang et~al.(2025)H.~Wang, J.~Lin, B.~Chen, Y.~Yang, R.~Tang, W.~Zhang, Y.~Yu]{wang2025towards}
{\EM Wang Hangyu, Lin Jianghao, Chen Bo, Yang Yang, Tang Ruiming, Zhang Weinan, Yu~Yong}.
\newblock Towards efficient and effective unlearning of large language models for recommendation \allowbreak\newblock// Frontiers of Computer Science. 2025. 19, 3. 193327.

\bibitem[Wei et~al.(2022)J.~Wei, X.~Wang, D.~Schuurmans, M.~Bosma, F.~Xia, E.~Chi, Q.~V. Le, D.~Zhou, et~al.]{wei2022chain}
{\EM Wei Jason, Wang Xuezhi, Schuurmans Dale, Bosma Maarten, Xia Fei, Chi Ed, Le~Quoc~V, Zhou Denny, others }.
\newblock Chain-of-thought prompting elicits reasoning in large language models \allowbreak\newblock// Advances in neural information processing systems. 2022. 35. 24824--24837.

\bibitem[Xi et~al.(2024)Y.~Xi, W.~Liu, J.~Lin, B.~Chen, R.~Tang, W.~Zhang, Y.~Yu]{xi2024memocrs}
{\EM Xi~Yunjia, Liu Weiwen, Lin Jianghao, Chen Bo, Tang Ruiming, Zhang Weinan, Yu~Yong}.
\newblock MemoCRS: Memory-enhanced Sequential Conversational Recommender Systems with Large Language Models \allowbreak\newblock// Proceedings of the 33rd ACM International Conference on Information and Knowledge Management. 2024.  2585--2595.

\bibitem[Xiong et~al.(2020)L.~Xiong, C.~Xiong, Y.~Li, K.-F. Tang, J.~Liu, P.~Bennett, J.~Ahmed, A.~Overwijk]{xiong2020approximate}
{\EM Xiong Lee, Xiong Chenyan, Li~Ye, Tang Kwok-Fung, Liu Jialin, Bennett Paul, Ahmed Junaid, Overwijk Arnold}.
\newblock Approximate nearest neighbor negative contrastive learning for dense text retrieval \allowbreak\newblock// arXiv preprint arXiv:2007.00808. 2020.

\bibitem[Xu et~al.(2023)Q.~Xu, F.~Hong, B.~Li, C.~Hu, Z.~Chen, J.~Zhang]{xu2023tool}
{\EM Xu~Qiantong, Hong Fenglu, Li~Bo, Hu~Changran, Chen Zhengyu, Zhang Jian}.
\newblock On the tool manipulation capability of open-source large language models \allowbreak\newblock// arXiv preprint arXiv:2305.16504. 2023.

\bibitem[Yao et~al.(2022)S.~Yao, J.~Zhao, D.~Yu, N.~Du, I.~Shafran, K.~Narasimhan, Y.~Cao]{yao2022react}
{\EM Yao Shunyu, Zhao Jeffrey, Yu~Dian, Du~Nan, Shafran Izhak, Narasimhan Karthik, Cao Yuan}.
\newblock React: Synergizing reasoning and acting in language models \allowbreak\newblock// arXiv preprint arXiv:2210.03629. 2022.

\bibitem[Yuan et~al.(2024)S.~Yuan, K.~Song, J.~Chen, X.~Tan, Y.~Shen, R.~Kan, D.~Li, D.~Yang]{yuan2024easytool}
{\EM Yuan Siyu, Song Kaitao, Chen Jiangjie, Tan Xu, Shen Yongliang, Kan Ren, Li~Dongsheng, Yang Deqing}.
\newblock Easytool: Enhancing llm-based agents with concise tool instruction \allowbreak\newblock// arXiv preprint arXiv:2401.06201. 2024.

\bibitem[Zhang et~al.(2024)Y.~Zhang, X.~Fan, J.~Wang, C.~Chen, F.~Mo, T.~Sakai, H.~Yamana]{zhang2024data}
{\EM Zhang Yuxiang, Fan Xin, Wang Junjie, Chen Chongxian, Mo~Fan, Sakai Tetsuya, Yamana Hayato}.
\newblock Data-Efficient Massive Tool Retrieval: A Reinforcement Learning Approach for Query-Tool Alignment with Language Models \allowbreak\newblock// Proceedings of the 2024 Annual International ACM SIGIR Conference on Research and Development in Information Retrieval in the Asia Pacific Region. 2024.  226--235.

\bibitem[Zheng et~al.(2023)L.~Zheng, W.-L. Chiang, Y.~Sheng, S.~Zhuang, Z.~Wu, Y.~Zhuang, Z.~Lin, Z.~Li, D.~Li, E.~P. Xing, H.~Zhang, J.~E. Gonzalez, I.~Stoica]{zheng2023judging}
{\EM Zheng Lianmin, Chiang Wei-Lin, Sheng Ying, Zhuang Siyuan, Wu~Zhanghao, Zhuang Yonghao, Lin Zi, Li~Zhuohan, Li~Dacheng, Xing Eric.~P, Zhang Hao, Gonzalez Joseph~E., Stoica Ion}.
\newblock Judging LLM-as-a-judge with MT-Bench and Chatbot Arena. 2023.

\bibitem[Zheng et~al.(2024)Y.~Zheng, P.~Li, W.~Liu, Y.~Liu, J.~Luan, B.~Wang]{zheng2024toolrerank}
{\EM Zheng Yuanhang, Li~Peng, Liu Wei, Liu Yang, Luan Jian, Wang Bin}.
\newblock ToolRerank: Adaptive and Hierarchy-Aware Reranking for Tool Retrieval \allowbreak\newblock// arXiv preprint arXiv:2403.06551. 2024.

\bibitem[Zhu et~al.(2025)J.~Zhu, C.~Zheng, J.~Lin, K.~Du, Y.~Wen, Y.~Yu, J.~Wang, W.~Zhang]{zhu2025retrieval}
{\EM Zhu Jiachen, Zheng Congmin, Lin Jianghao, Du~Kounianhua, Wen Ying, Yu~Yong, Wang Jun, Zhang Weinan}.
\newblock Retrieval-Augmented Process Reward Model for Generalizable Mathematical Reasoning \allowbreak\newblock// arXiv preprint arXiv:2502.14361. 2025.

\end{thebibliography}
}

\newpage
\appendix

\definecolor{d_green}{RGB}{39,147,39}
\definecolor{d_yellow}{RGB}{255,173,23}

\section{Broader Impact}
\label{app: Broader}
Our work contributes to improving the quality of tool use in large language models and lays a foundation for enhancing the fundamental tool-use capabilities of large language models, especially when facing the web-scale massive amount of tools available for complex user queries. 
However, the training data required may involve personal information, potentially raising concerns about privacy and security.

\section{Methodology Details}

\subsection{LightGCN Operations for QC-GCN}
\label{app:QC-GCN}
We give the detailed computational process of applying LightGCN within QC-GCN to obtain the final graph representations, as introduced in Section~\ref{sec:QC-GCN}. Here, we take the query-tool graph as an example, while the operations on the query-scene graph are the same.

Specifically, we pass the query \(q\) and tool  \(t\) through the pretrained language model (PLM) to get the initial query and tool embeddings, and further obtain the scene embeddings via the average pooling over tool embeddings from the golden tool set  \(\mathcal{T}_q\):
\begin{equation}
\begin{aligned}
    e_q^{0}=\operatorname{PLM}(q), \;\; e_t^{0}=\operatorname{PLM}(t), \;\; e_s^{0}=\frac{1}{|\mathcal{T}_q|}\sum\nolimits_{t\in\mathcal{T}_q}e_t^{0}.
\end{aligned}
\end{equation}
Then, we apply multi-layer message propagation on the two graphs. 
Since the operations for the two graphs are symmetric, we take the query-tool graph as an example, and similar operations can be conducted for the query-scene graph. 
At the $l$-th layer for the query-tool graph, the message propagation can be written as:
\begin{equation}
\begin{aligned}    
    e_q^{l} = \sum_{t\in\mathcal{N}_{q}}\frac{e_t^{l-1}}{\sqrt{|\mathcal{N}_q|}\sqrt{|\mathcal{N}_t|}}, \; e_t^{l} = \sum_{q\in\mathcal{N}_{t}}\frac{e_q^{l-1}}{\sqrt{|\mathcal{N}_q|}\sqrt{|\mathcal{N}_t|}},\, l=1,\dots,L,
\end{aligned}
\end{equation}
where $\mathcal{N}_q$ and $\mathcal{N}_t$ are the neighbor sets for query $q$ and tool $t$, respectively. 
Next, we sum up the outputs from each layer to obtain the final representations for the query-tool graph:
\begin{equation}
    e_q^{QT}=\sum\nolimits_{l=0}^L e_q^{l},\;\;e_t^{QT}=\sum\nolimits_{l=0}^L e_t^{l}.
\end{equation}

\subsection{Transfer Function Variants}
\label{app:transfer}
we provide the following alternative transfer functions in addition to the element-wise gating mechanism in MassTool:
\begin{itemize}[leftmargin=10pt]
    \item \textbf{Attention}. We apply self-attention~\cite{vaswani2017attention} over $h$ and $e_q^{search}$, and perform element-wise addition to fuse attentive output vectors.
    \item \textbf{Concatenation}. We concatenate $h$ and $e_q^{search}$, and feed them to a multi-layer perceptron (MLP) for implicit knowledge fusion. 
    \item \textbf{Addition}. We conduct element-wise addition for $h$ and $e_q^{search}$. This transfer function is simple but not learnable.
\end{itemize}

\section{Experiment Setup Details}

\subsection{Datasets Statistics}
\label{app:dataset statistics}

We give the detailed dataset statistics in Table~\ref{tab:datasets_Statistics}, including the three benchmarking datasets and our developed ToolDet dataset.
Datasets are split into training and testing sets with a ratio of 9:1. 

\begin{table}[h]
\centering
\caption{Statistics of the datasets.}
\label{tab:datasets_Statistics}
\resizebox{0.66\textwidth}{!}{
\renewcommand\arraystretch{1.0}
\begin{tabular}{lcccccc}
\toprule
\multirow{2}{*}{\textbf{Dataset}} & \multicolumn{3}{c}{\textbf{\# Query}} & \multirow{2}{*}{\textbf{\# Tool}} & \multirow{2}{*}{
\begin{tabular}[c]{@{}c@{}} \textbf{\# Golden Tools}\\ \textbf{per Query} \end{tabular}
}\\
\cmidrule(lr){2-4}
 & \textbf{Train} & \textbf{Test} & \textbf{Total} & &\\
\midrule
ToolLens & 16,893 & 1,877 & 18,770 & 464 & 1 $\sim$ 3 \\
ToolBenchG2 & 74,257 & 8,250 & 82,507 & 11,473 & 2 $\sim$ 4 \\
ToolBenchG3 & 21,361 & 2,373 & 23,734 & 1,419 & 2 $\sim$ 4 \\
ToolDet (Ours) & 28,106 & 3,123 & 31,229 & - & - \\
\bottomrule
\end{tabular}
}
\end{table}

\subsection{Baselines}
\label{app: Baselines}
We select four model-agnostic tool retrieval methods as our baselines: 
\begin{itemize}
    \item \textbf{QTA}~\citep{zhang2024data} uses LLMs to rewrite user queries to further enrich the training data for tool retrievers. 
    \item \textbf{MMRR}~\citep{kachuee2024improving} employs LLMs to generate synthetic tool descriptions based on the user input query, and then retrieves similar tools from the global pool. 
    \item \textbf{APIRetriever}~\citep{qin2023toolllm} finetunes the neural retriever with refined query logs and tool documents. 
    \item \textbf{COLT}~\citep{qu2024colt} designs cross-view graph learning for better collaborative patterns among tools.
\end{itemize}

\subsection{Implementation Details}
\label{app:implementation}
We utilize the BEIR~\citep{thakur2021beirheterogenousbenchmarkzeroshot} framework to implement dense retrieval backbone models. 
We freeze the dense retrieval backbone models during the training of MassTool. 
We adopt Adam optimizer with a batch size of 2048, and select the learning rate from the set \{1e-3, 5e–3, 1e–4, 5e–4, 1e–5\}. 
For SUIM, the number of neighbors $K$ is chosen from $\{10, 20, 25, 30, 35, 40\}$, and the threshold $\epsilon$ for dynamic filtering is selected from $\{0.6,0.7,0.8,0.9,1.0\}$. 
We set the contrastive loss weight $\beta$ as $0.04$, and select the tool usage detection loss weight $\lambda$ from $\{0.08,0.2,0.5,1.0\}$.
For fair comparison, we finetune all
the baselines to achieve their best performance.

\subsection{Ablation settings}
\label{app:ablation}
We conduct the ablation study by evaluating the performance of the following model variants:

\begin{itemize}[leftmargin=10pt]
    \item \textbf{MassTool}. The complete version of our proposed framework.
    \item \textbf{MassTool w/o AdaKT}. We remove the adaptive knowledge transfer module, resulting in single-tower learning for tool retrieval task without tool usage detection.
    \item \textbf{MassTool w/o SUIM}. We remove the search-based user intent modeling, and AdaKT directly takes as input the graph-enhanced query representation instead of the search-enhanced one.
    \item \textbf{MassTool w/o DF}. We remove the dynamic filtering mechanism (DF) in SUIM, and therefore the number of nearest neighbors for every input query stays the same, \ie, $K$.
    \item \textbf{MassTool w/o CL}. We remove the contrastive loss in Eq.~\ref{eq:contrastive loss}. 
\end{itemize}

\section{Additional Experiments}
\label{app:addtional exp}

\subsection{Hyperparameter Study}

\begin{figure}[t]
  \centering
  \includegraphics[width=\textwidth]{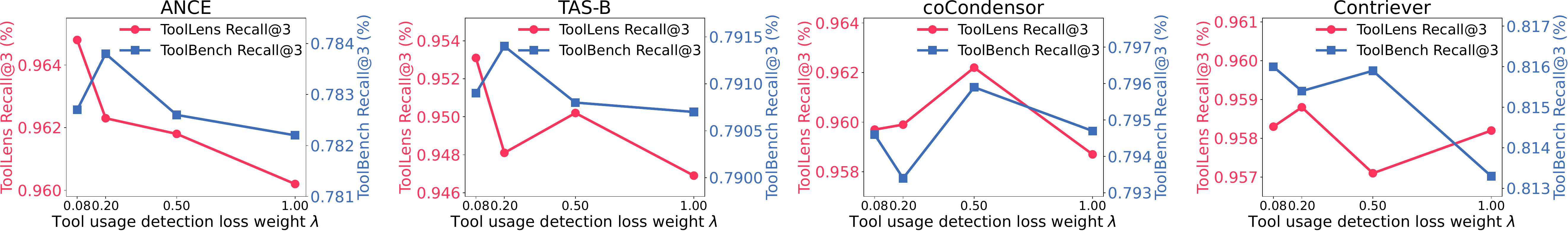}
  \caption{The performance of MassTool w.r.t. different tool usage detection loss weight $\lambda$ for the combined loss objective on ToolLens (\textcolor{red}{red} line) and ToolBenchG3 (\textcolor{blue}{blue} line) datasets.
  }
  \label{fig:aware}
\end{figure}

\paragraph{Tool usage detection loss weight $\lambda$.}
We select the weight of tool usage detection loss $\lambda$ from \{0.08, 0.2, 0.5, 1.0\}, and report the results in Figure~\ref{fig:aware}. 
We can observe that MassTool generally favors a relatively small loss weight for tool usage detection, compared with the tool retrieval loss weight which is set to 1. 
The binary tool usage detection task is significantly simpler than the tool retrieval task. 
Therefore, we need to assign a smaller loss weight to the detection task to slow down its learning pace. 
This ensures that the learning progress and convergence speed of two towers in the MassTool model remain synchronized, preventing the tool usage detection tower from overfitting prematurely.

\begin{figure}[t]
  \centering
  \includegraphics[width=\textwidth]{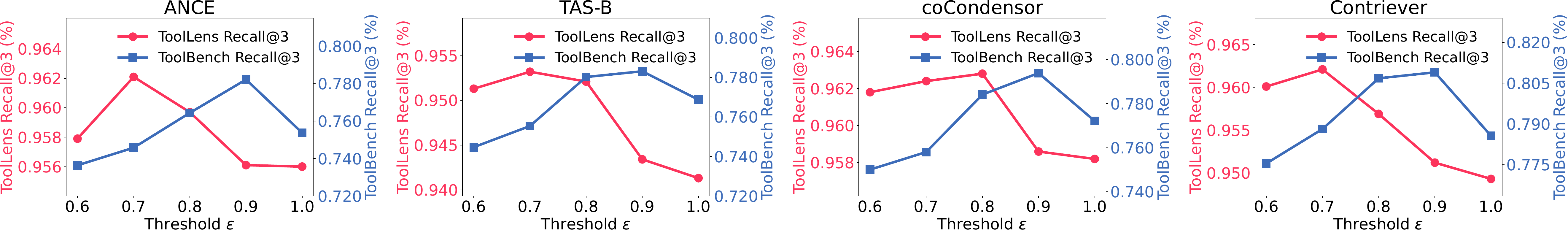}
  \caption{The performance of MassTool w.r.t. different thresholds $\epsilon$ for dynamic filtering in search-based user intent modeling (SUIM) on ToolLens (\textcolor{red}{red} line) and ToolBenchG3 (\textcolor{blue}{blue} line) datasets.
  }
  \label{fig:threshold}
\end{figure}
\paragraph{Threshold $\epsilon$ for dynamic filtering.}

We study the impact of the threshold $\epsilon$ for dynamic filtering in SUIM. 
We select the threshold $\epsilon$ from $\{0.6, 0.7, 0.8, 0.9, 1.0\}$ and report the results on ToolLens and ToolBenchG3 datasets in Figure~\ref{fig:threshold}. 
Note that $\epsilon=1.0$ means that we remove the dynamic filtering and accept all the top-$K$ nearest neighbors. 
We observe that the optimal value for threshold generally lies in the range of $[0.7, 0.9]$, indicating the importance of eliminating potential noisy neighbor queries.

\subsection{Query Representation Visualization}

\begin{figure}[h]
  \centering
  \includegraphics[width=0.7\textwidth]{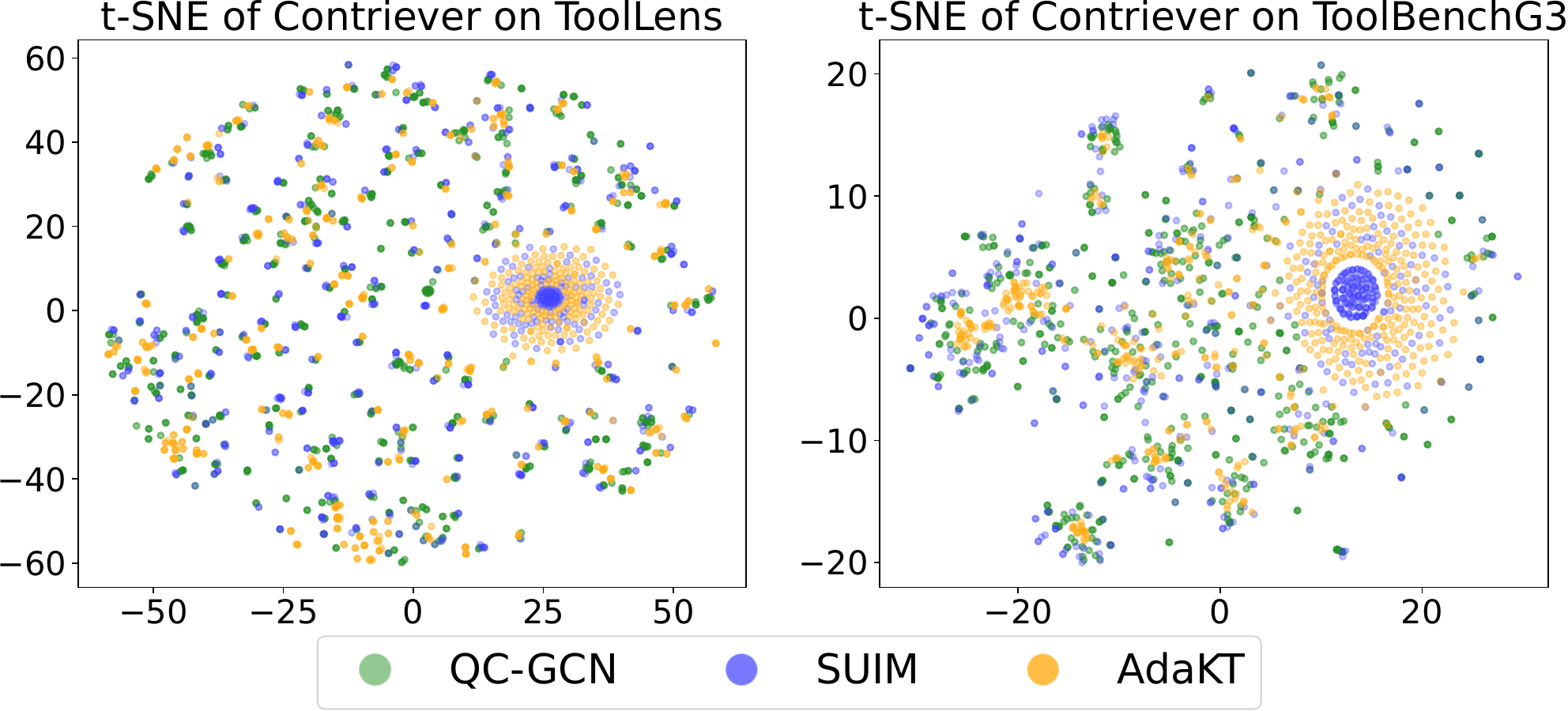}
  \caption{
  The t-SNE visualization for query representations from QC-GCN (in \textcolor{d_green}{green}), SUIM (in \textcolor{blue}{blue}), AdaKT (in \textcolor{d_yellow}{yellow}). 
  }
  \label{fig:query visualization}
\end{figure}

To showcase the capability of MassTool in modeling multi-aspect information of input queries for tool retrieval, we visualize the different sets of query representations in MassTool using the t-SNE projection~\cite{van2008visualizing}. 
Specifically, we visualize the following three different types of query representations based on the Contriever backbone model on ToolLens and ToolBenchG3 datasets:

\begin{itemize}[leftmargin=10pt]
    \item Graph-enhanced query representations $e_q^{graph}$ from query-centric graph convolution network (QC-GCN).
    \item Search-enhanced query representations $e_q^{search}$ from search-based user intent modeling (SUIM).
    \item Jointly fused query representations $e_q^{joint}$ from adaptive knowledge transfer module (AdaKT).
\end{itemize}
We uniformly sample 300 queries from the tool retrieval dataset $\mathcal{D}_{ret}$, and show the visualization in Figure~\ref{fig:query visualization}. 
We can observe that graph-enhanced query representations (in green) generally cast a uniform distribution, while the search-enhanced (in blue) and jointly fused (in yellow) query representations show more compact but different distributional manifolds within the query embedding space. 
We argue that QC-GCN is designed to capture overall and basic user intents based on the query-tool-scene collaborative patterns, and SUIM and AdaKT further refine and specify on more diverse, rare and nuanced aspects of the input query. 
As a result, we can deeply explore the underlying user intent and thereby optimize the query representations for better tool retrieval performance.

\end{document}